\documentclass[
      aps,
      prb,
      twocolumn,
      final,
      superscriptaddress,
      showpacs, 
      floatfix,
      longbibliography ]{revtex4-2}
      
\usepackage{bm}					

\usepackage{graphicx}

\usepackage{amsmath,amsfonts,amssymb,mathtools}	
\usepackage{latexsym} 				
\usepackage[usenames,dvipsnames, pdftex]{xcolor}
\usepackage[colorlinks, breaklinks, linkcolor=OrangeRed,citecolor=RoyalBlue, urlcolor=RoyalBlue]{hyperref}
\usepackage{braket}
\usepackage{blindtext}
\usepackage{txfonts}
\usepackage{soul}
\usepackage{xspace} 

\graphicspath{{images/}}

\newcommand{\be}{\begin{equation}}
\newcommand{\ee}{\end{equation}}
\newcommand{\bea}{\begin{eqnarray}}
\newcommand{\eea}{\end{eqnarray}}

\newcommand{\br}{{\bf r}}

\newcommand{\rmx}{\zeta}

\newcommand{\Nsamples}{N_\text{\tiny samples}}
\newcommand{\qplus}{q_{+}^{*}} 
\newcommand{\Pss}{P^\mathrm{(SS)}}

 \def\stacksymbols #1#2#3#4{\def\theguybelow{#2}
    \def\verticalposition{\lower#3pt}
    \def\spacingwithinsymbol{\baselineskip0pt\lineskip#4pt}
    \mathrel{\mathpalette\intermediary#1}}
\def\intermediary#1#2{\verticalposition\vbox{\spacingwithinsymbol
      \everycr={}\tabskip0pt
      \halign{$\mathsurround0pt#1\hfil##\hfil$\crcr#2\crcr
               \theguybelow\crcr}}}

\begin{document}
\title{Quartic multifractality and finite-size corrections at the spin quantum Hall transition}

\author{Martin Puschmann} 
\affiliation{Institute of Theoretical Physics, University of Regensburg, D-93053 Regensburg, Germany}

\author{Daniel \surname{Hernang\'{o}mez-P\'{e}rez}}

\affiliation{Department of Molecular Chemistry and Material Science, Weizmann Institute of Science, Rehovot 7610001, Israel}

\author{Bruno Lang}

\affiliation{IMACM and Institute of Applied Computer Science, Bergische Universit\"at Wuppertal, D-42119 Wuppertal, Germany}

\author{Soumya Bera}

\affiliation{Department of Physics, Indian Institute of Technology Bombay, Mumbai 400076, India}

\author{Ferdinand Evers}

\affiliation{Institute of Theoretical Physics, University of Regensburg, D-93053 Regensburg, Germany}

\date{\today} 
\begin{abstract}
The spin quantum Hall transition (or class C transition in two dimensions) represents one of the few localization-delocalization transitions for which some of the critical exponents are known exactly. Not known, however, is the multifractal spectrum, $\tau_q$, which describes the system-size scaling
of inverse participation ratios $P_q$, i.e., the $q$-moments of critical wavefunction amplitudes. We here report simulations based on the class C Chalker-Coddington network and demonstrate that $\tau_q$ is (essentially) a quartic  polynomial in $q$. Analytical results fix all prefactors except the quartic curvature that we obtain as $\gamma=(2.22\pm{0.15})\cdot10^{-3}$. In order to achieve the necessary accuracy in the presence of sizable corrections to scaling, we have analyzed the evolution with system size of the entire $P_q$-distribution function. As it turns out, in a sizable window of $q$-values this distribution function exhibits a (single-parameter) scaling collapse already in the pre-asymptotic regime, where finite-size corrections are {\em not} negligible. This observation motivates us to propose a novel approach for extracting $\tau_q$ based on concepts borrowed from the Kolmogorov-Smirnov test of mathematical statistics. We believe that our work provides the conceptual means for high-precision investigations of multifractal spectra also near other localization-delocalization transitions of current interest, especially the integer (class A) quantum Hall effect. 
\end{abstract}
\maketitle
\section{Introduction}
The Altland-Zirnbauer classification of disordered metals exhibits several symmetry 
classes that allow for insulating and metallic phases, which are separated 
by critical points or critical lines. Prominent examples include the integer, spin and thermal 
quantum Hall effects in classes A, C, D or the Anderson transition 
in the symplectic class AII \cite{Evers2008RMP}.
A distinguishing feature of these transitions 
is the statistics of the amplitudes of critical wavefunctions, $\psi(\br)$, 
which is conveniently described by their moments, 
\textit{i.e.} the inverse participation ratios (IPR) $P_q(L)= \int d\br  |\psi(\br)|^{2q}$.
For critical wavefunctions, the scaling of the average IPR with the system size, $L$, 
\be
\overline{P_q}(L) \sim L^{-\tau_q},
\ee
defines the set of multifractal exponents $\tau_q$; the overline indicates a combined averaged over a narrow spectral window and an ensemble of disorder 
configurations (``samples").

The $\tau_q$-spectrum is of interest because it is a characteristic fingerprint of a critical state and the corresponding field theory. 
For instance, a long-standing conjecture by Zirnbauer \cite{Zirnbauer1999} for the critical theory of the class A transition, which has undergone a recent refinement \cite{Zirnbauer2017,Zirnbauer2019}, 
implies that the multifractal spectrum takes a particularly simple form: 
$\tau_q$ is exactly parabolic in $q$.
Therefore, the multifractal spectrum is a standard means to compare field theoretical predictions and numerical simulations for microscopic models~\cite{Evers2008RMP}. 


An important recent field-theoretic result~\cite{Gruzberg2007, Gruzberg2011, Mirlin2006} indicates that the spectrum $\tau_q$ for a transition in $d$-dimensions has an exact representation 
\be
\tau_q = d(q-1) - qx_1 + x_q,
\ee
in terms of the exponents $x_q$ that describe the scaling of the local 
density of states at criticality 
\be
\overline{ \rho^q(\mathbf{r})}\sim L^{-x_q}.  \ee
Crucially, under very general conditions the exponents $x_q$ obey a reciprocity relation~\cite{Mirlin2006,Gruzberg2011,Gruzberg2013}  
\be
\label{e5}
x_q = x_{q^*-q},
\ee
with $q^\ast$ taking different values for different symmetry classes, \textit{e.g.}, 
$q^\ast=1$ for class A and $q^\ast=3$ for class C~\cite{Gruzberg2011}. 
Even more intriguing is the result of Ref.~\cite{Mirlin1994,Fyodorov2004,Fyodorov2005,Gruzberg2011} according to which reciprocity is a property not just of the exponents but of the entire correlator itself: 
\be
\label{e6} 
\overline{\rho^q(\mathbf{r})} = 
\overline{\rho^{q^*-q}(\mathbf{r})},
\ee
in localized, delocalized and critical regimes. This remarkable statement reflects a Weyl-symmetry that is a property of the corresponding $\sigma$-model descriptions.  
\cite{Gruzberg2011} 

Inspired by the recent conjecture~\cite{Zirnbauer2017,Zirnbauer2019}, 
in this article we address the question of whether parabolicity of the multifractal spectrum could be a generic feature of quantum Hall transitions. Rather than investigating the class A transition considered by Zirnbauer {\it et al.}~\cite{Zirnbauer2017,Zirnbauer2019}, 
we here  focus on class C, \textit{i.e.} the spin quantum Hall transition. Here, several open questions need to be addressed:

(a) Taken at face value, the earlier numerical work on the class-C transition suggests pronounced deviations from parabolicity in the anomalous exponents $\Delta_q {=} \tau_q - d(q-1)$~\cite{Mirlin2003,Evers2008RMP}. 
At present, it cannot be safely excluded that the observed deviations are artifacts due to finite-size effects that have not been included in the earlier data analysis. 
Specifically, we have exact results for the local density of states~(LDoS) exponent  $x_1=1/4$ and for the density correlator $x_2=1/4$~\cite{Gruzberg1999, Mirlin2003}. Also at $q=3$ an exact statement has been obtained corresponding to $x_3=0$~\cite{Mirlin2003}. While all these results are consistent with the reciprocity condition $x_q=x_{3-q}$ and a putative parabolic form  $x_q^\text{(p)}\coloneqq \frac{1}{8}q(3-q)$, they are not sufficient to rule out quartic (and higher) terms in $q(q-3)$. 

(b) The validity of the reciprocity relation has not yet been established numerically for class C. 

(c) Further, we also address aspects of class-C criticality  that have received less attention in the past, most notably the set of tail-exponents $\rmx_q$ that describes rare fluctuations with extremal IPR values. 

There is one more consideration of a methodological kind that motivates us to scrutinize the class-C transition once again. 
In contrast to class A, at the class-C transition two non-trivial exponents are known analytically. This is very helpful for the analysis of simulation data, because it gives reference points with respect to which finite-size effects can safely be quantified. We thus may use class-C  criticality as a laboratory to test known and explore novel approaches for fitting of critical exponents in the presence of sizable finite-size effects. In particular with an eye on open questions, e.g., in connection with the class-A transition, such methodological advancements would  certainly be welcome. 

We briefly list our most important results obtained from simulations within the framework of the class-C  Chalker-Coddington network. \cite{Gruzberg1999, Kagalovsky1999} 

(i) The reciprocity relation, $x_q=x_{3-q}$, 
as well as the analytically predicted exponents are reproduced with excellent accuracy: 
$x_2 = 0.2504 \pm 0.0008$; $x_3 = 0.002 \pm 0.005$ (within an accuracy of $0.2\%$).

(ii) Quartic (and possibly higher) deviations from a parabolic shape  
$x^\text{(p)}_q= q(3-q)/8$ exist,
\be
x_q = x^\text{(p)}[1+8\gamma_q(q-1)(q-2)],\nonumber
\ee
with $8\gamma_q\approx 0.0178\pm0.0012$.

(iii) The tail exponents take values $\rmx_q \approx 2\rmx_{2} |q (q-1)|^{-1}, \rmx_{2}=4.5\pm 0.5$ for $q\in[0,3]$. 

(iv) Within the framework of the  $\sigma$-model, i.e. by virtue of Eq. \eqref{e6}, the ratios 
\begin{equation} 
r_q(L)\coloneqq L^{(d-x_1)(2q-q^*)}\frac{\overline{P_q}(L)}{\overline{P_{q^*-q}}(L)}, 
\label{e6a} 
\end{equation} 
are not expected to exhibit corrections to scaling. Our data does not conform with this expectation; it exhibits discernible finite-size corrections
$r_q(L) = r_q^{*}(1+r_q^{(1)}L^{-y} +\ldots) $
with $y\approx 1$. 
Such corrections reflect microscopic features of lattice models that the coarse-grained scale of the $\sigma-$model is ignorant of by design. 
In the present context they indicate that the $\sigma$-model's Weyl symmetry is an  emerging property. 

(v) Traditionally, the ensemble averaged IPR has received most of the attention, because it is an object that appears naturally in quantum field theories \cite{Wegner1980} where it corresponds to a well defined scaling operator. We here investigate the scaling properties of the entire IPR-distribution functions, ${\mathcal P}_q$. Because they are known to be more sensitive to finite-size effect, they lend themselves for sensing of these. Actually, we find that the distributions satisfy a non-trivial scaling relation
\begin{equation} 
{\mathcal P}_q(\Lambda_q;L)=\frac{\lambda^*_q}{\lambda_q}\
{\mathcal P}^\infty_q\left(\frac{\lambda_q^*}{\lambda_q}\{\Lambda_q-c_q\}+c_q \right),
\label{e6b} 
\end{equation} 
valid within its bulk region; here, 
$\Lambda_q(L)=\ln P_q L^{\tau_q}$, 
$\lambda_q^*,c_q$ constants and $\lambda_q(L)/\lambda_q^*=1+\mathcal{O}(L^{-y})$. 

(vi) We propose a methodological advancement in the analysis of finite-size effects on the IPR-scaling that relies on a standard parameter-free statistical test, the Kolmogorov-Smirnov test. \cite{DeGroot2012} It allows to extract $\tau_q$ from the flow of the entire IPR-distribution by fitting only a single parameter, even in the presence of significant finite-size correction. 

\section{Model and Methods}
{\subparagraph{Model and method.}
We consider the version of the  Chalker-Coddington network (CCN) \cite{Chalker1988} 
adapted to the spin quantum Hall transition (SQH) \cite{Kagalovsky1999, Senthil1999, Gruzberg1999, Beamond2002, Evers2003}. Since the model has been described extensively before \cite{Kagalovsky1999,Evers2008RMP}, 
we allow ourselves to be brief. 
The class C network model consists of scattering nodes, $\Sigma$, arranged on a square lattice of size $ L^2/2$ nodes. Neighboring nodes are connected by unidirectional links with opposing pairs incoming and outgoing. 
Every link carries two spin channels, $\sigma \in \{\pm\}$ that  mix while propagating along 
as prescribed by mixing matrices, $U_l$,  chosen at random from a $\textnormal{SU}(2)$ distribution
but fixed for every link $l$. 
%
Nodal scattering for an incoming link is left/right; it is incorporated by orthogonal 
scattering matrices, $S$, diagonal in spin space, $S^{\Sigma} = S_{+}^{\Sigma} \otimes S_{-}^{\Sigma}$. 
At criticality, scattering to the left and right occurs with the same probability, $ p = 1/2$.  
%

The network dynamics is mediated by the unitary network operator $\mathcal{U}$; 
it is the direct product of the mixing matrices, $U_l$, and the nodal scattering operators,
$S^\Sigma$, and it  describes the evolution of the set of link probability 
amplitudes $\{ \psi_{ l+}, \psi_{l-}\}$, $l=1,\ldots,L^2$ in discretized time. 
\cite{Klesse1995, Klesse1999}. 
The eigenvalues of $\mathcal U$ appear in 
pairs, $\exp(\pm\mathfrak{i}\epsilon)$.
We construct $\mathcal{U}$ for a given disorder realization
and use a variant of the Lanczos algorithm to extract six eigenstates, $|\Psi\rangle$, 
corresponding to the three pairs of eigenvalues closest to unity. 
For the statistical analysis we employ one of the eigenvectors with eigenvalue closest to unity; the remaining two eigenstates are used as consistency check, see Appendix\;\ref{aC}.
We simulated networks with linear system sizes
$L = 16, 24, 32, \dots, 1024$ on a toroidal geometry and considered ${\sim}10^7$ samples for the smallest and 
${\sim}10^6$ samples for the largest system size, see App. \ref{aA} for more details.  
\subparagraph{Observable.} In lattice or network models there can be different microscopic observables that all flow towards the same macroscopic object after coarse graining. A typical example is the IPR in the CCN model, where the local weight per link can be, e.g., 
either $|\psi_{+}|^2$ or $|\psi_{-}|^2$ or the sum of both. All these local weights give rise to different flavors of the IPR. The difference will become manifest when rare events are considered: roughly speaking, a link with a particularly low weight is a lot more probable if just one amplitude is assigned to the link. 

The weight we chose to implement is a physical observable, i.e. the occupation per link: 
\begin{equation}
   P_q(L) = \sum_l \left( \sum_\sigma |\psi_{l\sigma}|^2\right)^q, \label{e7}
 \end{equation}
 with conventional normalization $P_1(L)=1$. In the Appendix\;\ref{aD}, we consider other flavors of IPR and show that the different weighting does not affect asymptotic scaling properties.

\section{Selected pairs of $q$-values}

\subsection{Finite-size effects at $q=2$ and $q=3$}
We begin the presentation of our numerical results with the IPR at $q=2$ and $q=3$, because in these cases $\tau_q$ is known analytically: 
$\tau_2=7/4$ and $\tau_3=13/4$. Therefore, the irrelevant corrections to the scaling are readily exposed when considering {\it reduced}  IPRs, $P_j L^{\tau_j}$ with $j=2,3$. The understanding here achieved we then will transfer to the case of general $q$-values.

\subsubsection{The case $q=2$} 
\begin{figure}[t]
	\includegraphics{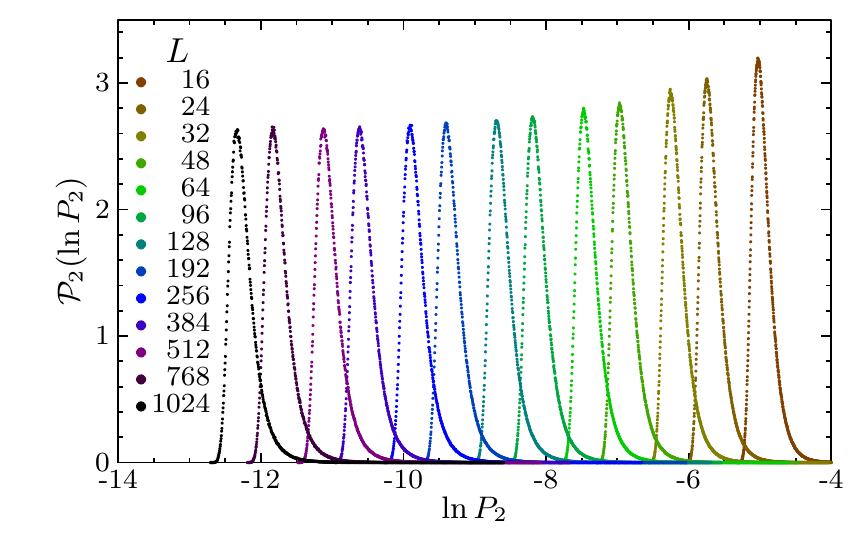}
\caption{Probability distribution ${\mathcal P}(\ln P_q,L)$ of the IPR for $q=2$, 
presented for several system sizes $L=16,..,1024$.
\label{fig1}}
\end{figure}
 Fig. \ref{fig1} shows the evolution (``flow") of the distribution function 
${\mathcal P}_2$
of $\ln P_2(L)$ with system sizes. 
The typical behavior is seen: the distribution becomes shape-invariant under shifting at largest system sizes. This shape-invariance is generally considered to emerge only very close to the critical point;  \cite{Mirlin2000} 
therefore, the flow of the shape itself can be considered an indicator of the closeness to criticality and therefore is of interest to us, here. 

 To monitor the  shape evolution we analyze the system-size dependency of its width 
 \begin{equation}
    \sigma_q(L)= \left[\overline{ \left(P_q(L)
    -\overline{P_q}(L)\right)^2}\right]^{1/2},
    \label{e9} 
\end{equation}
the result is displayed in Fig. \ref{fig2}. Another descriptor is the peak value $h_q(L)$ (``height") of the distribution, also shown in Fig.~\ref{fig2}. Due to statistical fluctuations, $h_q(L)$ was obtained 
from polynomial fits to the distribution dome, applied at each $L$ separately. The fitting interval was identified by comparing the mean deviation for cubic and quartic fits. Due to normalization, the system-size variation of the inverse width, $\sigma^{-1}_{q}(L)$, and the height $h_q(L)$ is identical in the limit of large $L$. Consequently, deviations from this behavior are indicative of pre-asymptotic changes in the shape of 
$\mathcal{P}_q$. 

The growth of $\sigma_2(L)$ with increasing $L$ in Fig. \ref{fig2} accounts for the enhanced variability of local wavefunction amplitudes, which in turn reflects the gradual unfolding of the localized character of the wavefunctions. 
\begin{figure}[t]
	\includegraphics{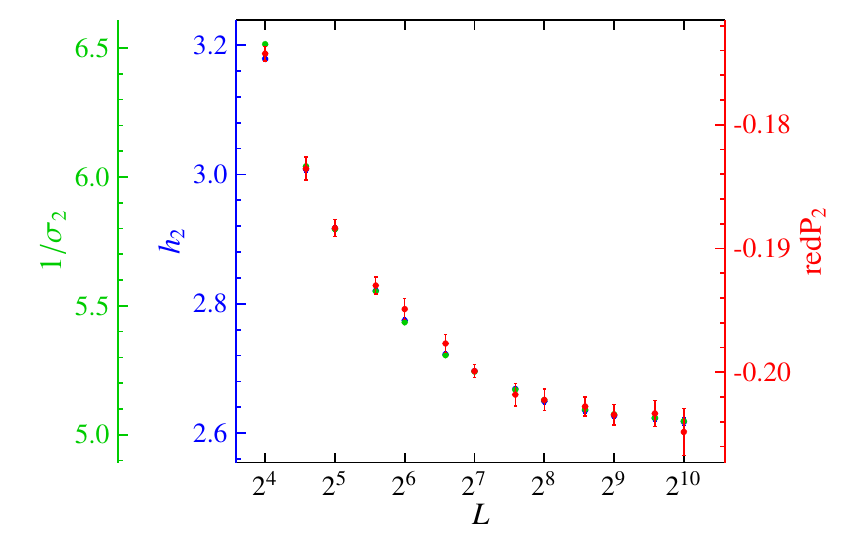}
	\caption{Evolution of the shape of the IPR-distribution ${\cal P}_2$ with increasing system size, $L$, monitored by its width $\sigma_{2}(L)$ and height $h_2(L)$. Error bars represent statistical uncertainties due to random fluctuations and fluctuations based on fit ranges. Also shown is the residual shift of the maximum of the distribution of $\ln L^{\tau_2}P_2$ (``reduced distribution"). \label{fig2}}
\end{figure}
\begin{figure}[b]
	\includegraphics{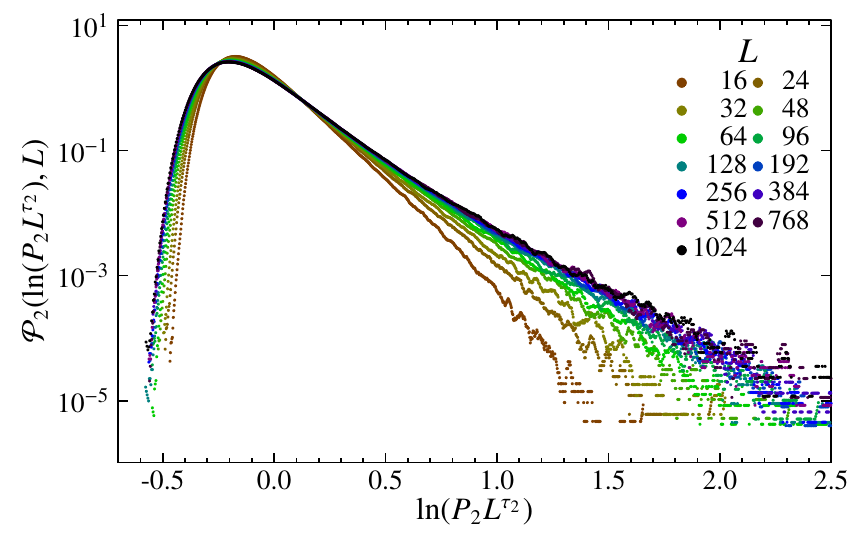}
	\caption{Probability distribution ${\mathcal P}_2(\ln P_2L^{\tau_2},L)$ (``reduced distribution"). At large system sizes a scaling collapse is observed indicating that the asymptotic scaling regime has been reached. 
	\label{fig3}}
\end{figure}
Similar finite-size corrections as seen in height and width also manifest in the position and, hence, in the flow of the average IPR, $\overline{P_2}(L)$.
For illustration we define the reduced peak position $\text{redP}_2(L)$, i.e. the position of the peak of the distribution of $\ln (L^{\tau_2} P_2)$. (The shifted argument implies subtracting the ``translation" with $L^{\tau_2}$.) 
For the reduced distribution 
${\cal P}_2(\ln (P_2 L^{\tau_{2}});L)$ a scaling collapse is expected in the limit $L\to\infty$ and indeed observed in Fig. \ref{fig3}.
The corresponding peak position $\text{redP}_2(L)$ as extracted from Fig. \ref{fig3} is also displayed in Fig. \ref{fig2}. Its saturation at large $L$ indicates that our data is indeed consistent with the expected theoretical exponent value.

A visual inspection of Fig. \ref{fig2}  suggests that the (inverse) width $1/\sigma_2(L)$ and the peak position $\text{redP}_2(L)$ exhibit a concerted flow towards criticality. 
A more quantitative analysis proceeds by stipulating a form 
\begin{eqnarray} 
\sigma_{q}(L)&=&\sigma^*_{q}\left(1+\sum_{j=1}^{N_y} \sigma^{(j)}_qL^{-jy}\ \right), \label{e8}
\end{eqnarray}
and similar for $h_q(L)$ and $\text{redP}_q(L)$. 
The specific form of the expansion
is motivated by two facts: First, the IPR is a pure scaling operator \cite{Gruzberg2013}, so that all observables deriving from it exhibit the same set of irrelevant exponents $y$. Second, we accommodate a single irrelevant scaling field, so only a single (irrelevant) exponent $y>0$ appears.
The details of the fitting procedure have been relegated to the appendix \ref{aB}. The extensive analysis yields two important conclusions: (i) good fits are obtained with $N_y=2$; these fits are stable, in particular, against variations in the raw-data set and with $N_y=1,3$. (ii) The irrelevant exponent is obtained with best accuracy in the window $q\in[0,2]$ where it takes values close to unity, $y=1.0\pm 0.2$.
\begin{figure}[t]
	\includegraphics[width=\linewidth]{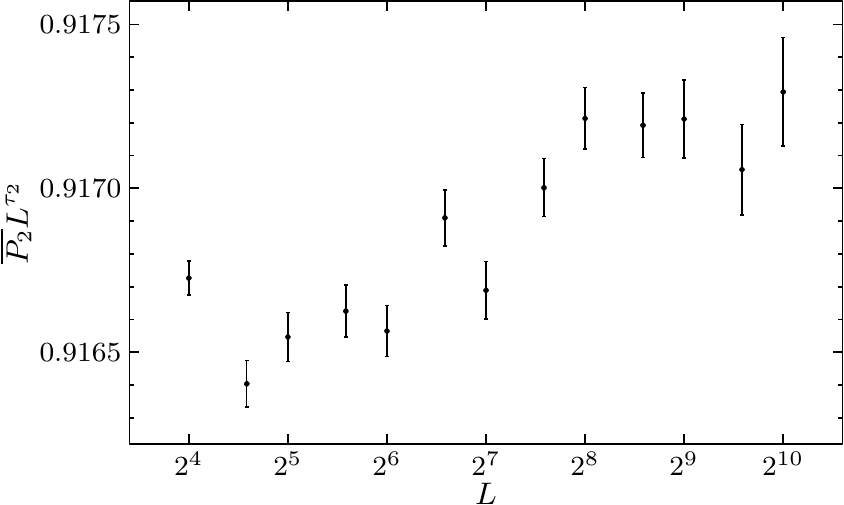}
	\caption{Corrections to scaling for the reduced moment $\overline{P_2} L^{\tau_2}$ highlighting finite-size effects. 
	\label{fig4}}
\end{figure}

Finite-size effects on the average IPR $\overline{P_2}(L)$ deserve a special attention. Remarkably, while the peak of the distribution function is seen to flow to the left in Fig. \ref{fig3}, an evolving power law tail strengthens the right hand side weight. 

Both effects cancel in the average $\overline{P_2}(L)$ to a surprising amount as seen in Fig. \ref{fig4}. 
It displays discernible, but weak finite-size corrections of the order of 0.1\% within our window of system sizes.

For better understanding, we discuss the presence of finite-size corrections in $\overline{P_2}(L)$ in the light of the reciprocity relation Eq. \eqref{e6}. 
At $q{=}2$ we have the special situation that
$L^{\tau_2}\overline{P_2}(L)=r_2(L)$, $\tau_2=7/4$,  since trivially $\overline{P_1}(L)=1$.
This is useful, because the reciprocity relation implies that at least within the framework of the 
$\sigma$-model the ratio $r_q(L)$ should not exhibit any scaling corrections, so also $L^{\tau_2}\overline{P_2}(L)$ is predicted to be independent of $L$.  %
In Fig. \ref{fig4}  corrections to scaling are seen, however; actually, from a microscopic perspective this is hardly surprising since power laws are not expected to hold in the limit where $L$ approaches the lattice constant.
 
We elaborate on this observation: The $\sigma$-model has a large (Weyl-type) symmetry; it eliminates corrections to scaling already at its short-distance cutoff, so that a field-theoretical perspective would predict a perfectly flat line in Fig. \ref{fig4}. We interpret our data deviating from flatness as an indication that the Weyl symmetry is, in principle, only approximate for microscopic models, such as the CCN. Consequently, lattice models can exhibit corrections to scaling also in observables that are correction-free on the $\sigma$-model level. Conversely, the fact that corrections in Fig. \ref{fig4} are seen to be so weak impressively illustrates how close the $\sigma$-model is to microscopic representations of the class-C transition. 

The general importance of expansions such as Eq. \eqref{e8} motivates one more remark. The corresponding expansion coefficients do not necessarily share the same sign. In fact we show in App. \ref{aB}, Figs. \ref{fig27} and \ref{fig28} for the specific example of $\sigma_q(L)$ that the first coefficient, $\sigma_q^{(1)}$, is likely negative, while the second one, $\sigma_q^{(2)}$, is positive. As a consequence, the effects of the first and second correction term partially cancel in a certain regime of system sizes ("conspiracy"), so that finite-size effects are very difficult to analyze. In the present situation this regime is narrow because $y{\approx}1$, i.e. rather large; therefore, conspiracy is less relevant for the class C transition. Since $y$ as reported\cite{Evers2008RMP,Obuse2013} for the class-A transition is much smaller, conspiracy is a more relevant issue for the integer quantum Hall effect.

\subsubsection{The case $q=3$} 
 \begin{figure}[t]
	\includegraphics{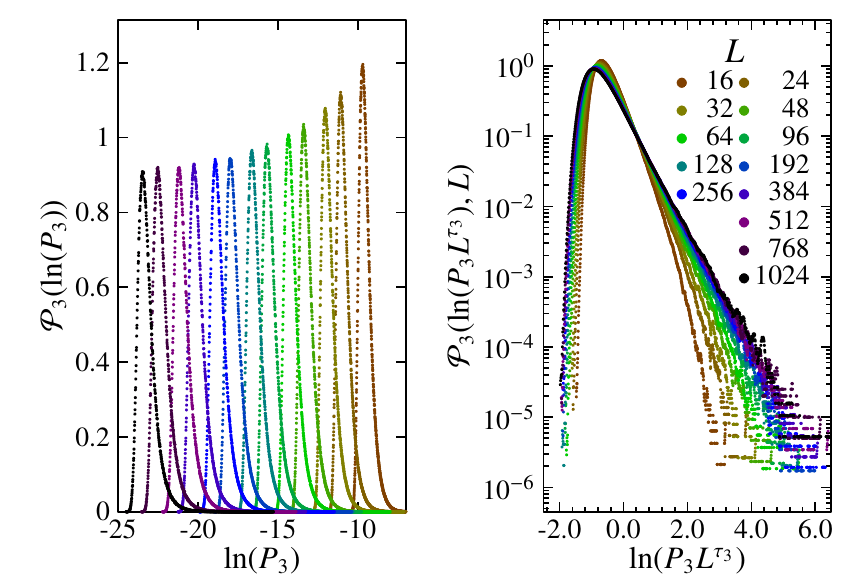}
	\caption{Data analogous to Fig. \ref{fig1} and Fig. \ref{fig3}, here for $q=3$. 
	\label{fig5}}
\end{figure}
\begin{figure}[b]
	\includegraphics{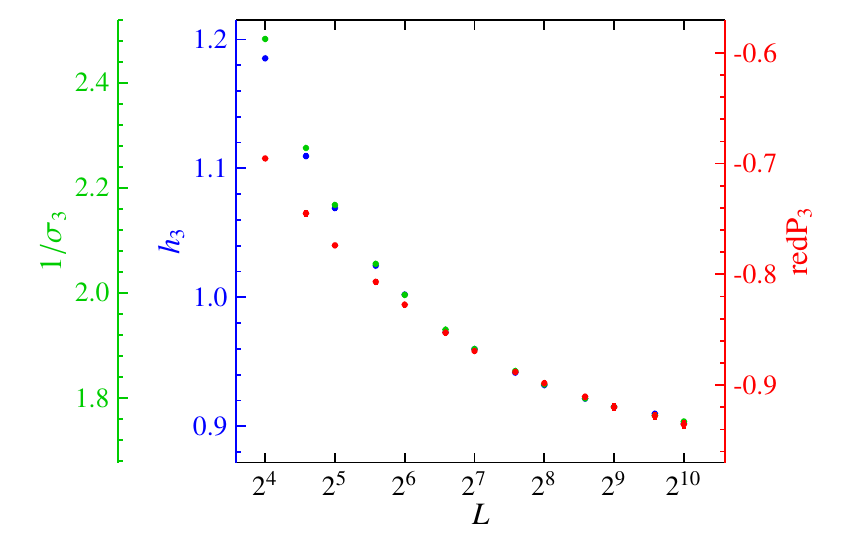}
	\caption{Data analogous to Fig. \ref{fig2}, here for $q=3$.\label{fig6}}
\end{figure}
For $q=3$ an analysis analogous to $q=2$ can be performed based on the data shown in Fig. \ref{fig5}. %
Also in this case the gradual evolution of ${\cal P}_3(\ln P_3 L^{\tau_3};L)$ terminates in a scaling collapse. 
The flow of the peak width, $\sigma_3(L)$, height $h_3(L)$ and peak position, $\text{redP}_3(L)$, is depicted in Fig.~\ref{fig6}. Its parametric analysis in terms of irrelevant corrections to scaling, Eq. \eqref{e8}, consolidates the picture developed above, see App. \ref{aB}. 
\begin{figure}[t]
	\includegraphics{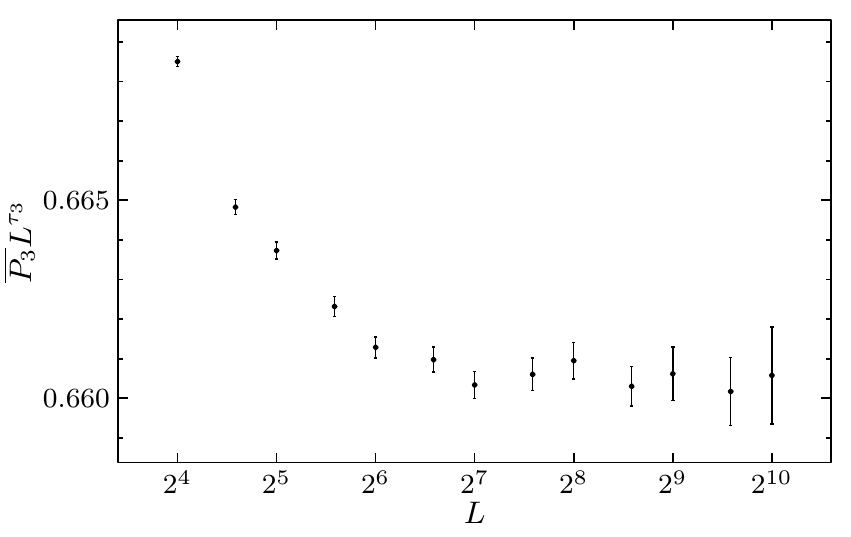}
	\caption{Plot analogous to Fig. \ref{fig4} highlighting finite-size effects in the reduced IPR, $\overline{P_3}L^{\tau_3}$, $\tau_3=13/4$. \label{fig7}}
\end{figure}

Also at $q=3$
reciprocity predicts that the scaling corrections for $L^{\tau_3+d} \overline{P_3}(L)$, $\tau_3+d{=}21/4$, vanish on level of the $\sigma-$model description.
We infer from Fig. \ref{fig7}, that the finite-size corrections seen in the microscopic model are pronounced, ten times larger as compared to the case $q=2$. The relative enhancement is not surprising:  higher moments, i.e. larger $q$-values, probe more extreme wavefunction amplitudes that are much more likely to occur in bigger systems. 

\subsection{Deviations from parabolic multifractality}
\subsubsection{General considerations} 
After presenting an analysis of the structure of the finite-size corrections at the class-C critical point, we now turn to the shape of the multifractal spectrum itself. Specifically, we will investigate potential deviations from a parabolic shape 
\begin{eqnarray}
\gamma_q&\coloneqq&\frac{\tau_q - \tau_q^\text{(p)}}{q(q-1)(q^*-q)(q-(q^*-1))}, \nonumber \\ &=& \frac{x_q - x_q^\text{(p)}}{q(q-1)(q^*-q)(q-(q^*-1))}, 
\label{e10} 
\end{eqnarray} 
with $\gamma_q = \gamma_{q^*-q}$. For class C,  $q^*=3$ and  $\tau_q^\text{(p)}\coloneqq d(q-1) - q/4 + q(3-q)/8$.  
The first two factors in the denominator are standard \cite{Evers2008RMP}; they accommodate the trivial zeros of the numerator. The second two factors reflect the reciprocity symmetry; they appear in those universality classes for which $q^*$ differs from unity.  Obviously, we have $\gamma=0$ for exact parabolicity.

It is implied by Eq. \eqref{e10} that 
    \begin{equation}
        x_q = \frac{1}{8}q(3-q)\left[1 + 8\gamma_q ( q-1) (q-2)\right], 
        \label{e12} 
    \end{equation}
while reciprocity symmetry suggests for the non-parabolicity parameter $\gamma_q$ an expansion of the form 
\begin{equation}
    \gamma_q =  \sum_{j=0} \gamma^{(j)} [(q-q^*/2)^2]^j. 
    \end{equation}
Stipulating a weak dependency of $\gamma_q$ on $q$,  Eq. \eqref{e12} suggests that  deviations from parabolicity are relatively small within the window $q\in[1,2]$ and more sizable outside. We therefore continue the analysis with the pair $q=1/2,5/2$. 

\subsubsection{The reciprocity pair $q=1/2$ and $5/2$} 
\begin{figure}[t]
	\includegraphics{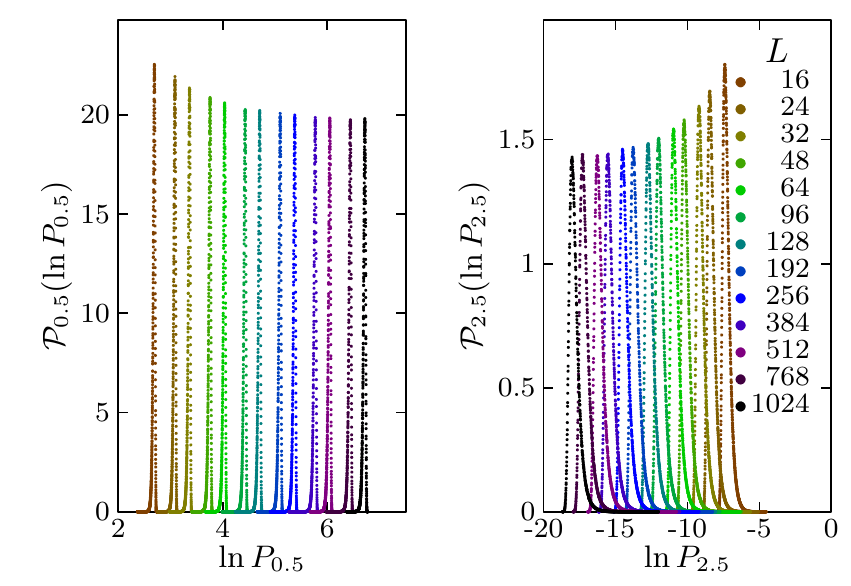}
	\caption{Probability distribution $\mathcal{P}_q(\ln P_q,L)$ of the IPR for $q=1/2$ (left) and $q=5/2$ (right). System sizes: $L=16,...,1024$.	\label{fig8}}
\end{figure}
As in the previous cases of $q=2$ and $q=3$ we begin with an analysis of  finite-size corrections visible in the raw data, see Fig. \ref{fig8}. Again we focus on a reduced IPR, which we now generalize to values away from $q=2,3$, so we investigate $L^{\tau_q^{(p)}}P_q(L)$. Our motivation is that in the presence of parabolicity one would expect a scaling collapse of $\mathcal{P}_q(L^{\tau_q^\text{(p)}}P_q(L))$. 
The flow of $\mathcal{P}_q$ of the reduced IPRs is parametrized in Fig. \ref{fig10}. 

The first information summarized in the three panels is that the reciprocity pairs at largest system sizes exhibit an identical scaling with respect to corrections to the system size. This is best illustrated for the evolution of $h_q(L)$, but also  $\sigma_q(L)$ is  eventually seen to follow this trend. We take this as an evidence that above $L\approx 2^8=256$ the finite-size corrections are dominated by a single correction term $L^{-y}$. 

The second information is that beyond this scale $L\approx 256$
 the reduced position $\text{redP}_q(L)$ does not show a tendency settling towards a horizontal line. We interpret this observation as evidence that deviations to parabolicity exist. 
\begin{figure}[t]
	\includegraphics[width=0.97\linewidth]{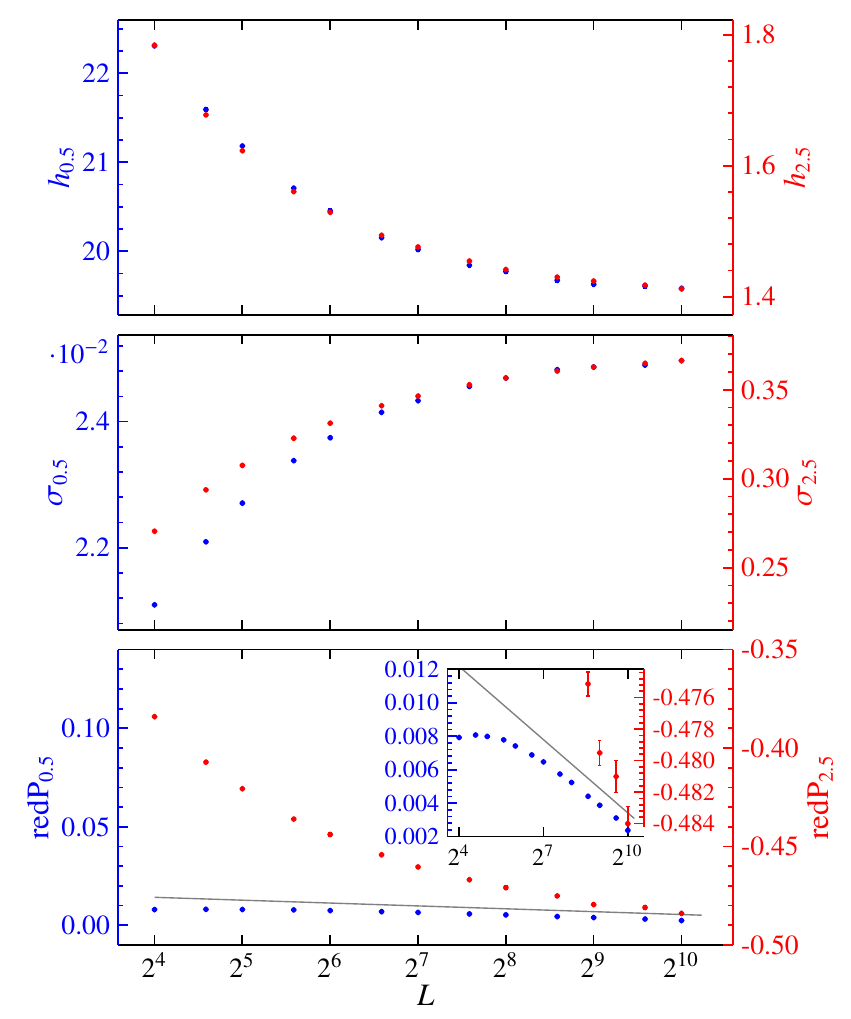}
	\caption{
	Convergence of the height $h_q(L)$ (top panel), the width $\sigma_q(L)$ (middle) and the reduced peak position $\text{redP}_q(L)$ (bottom) for $q=1/2$ and $5/2$ with system size $L$. The inset shows the asymptotic region in which the peak  positions shift in parallel for both $q$-values, consistent with  reciprocity. The solid line (slope corresponds to a $\gamma$-value 0.002244) is a guide to the eye indicating corrections to parabolicity in $\tau_q$ according to Eq. \eqref{e12}. \label{fig10}}
\end{figure}
\begin{figure}[b]
	\includegraphics[width=0.97\linewidth]{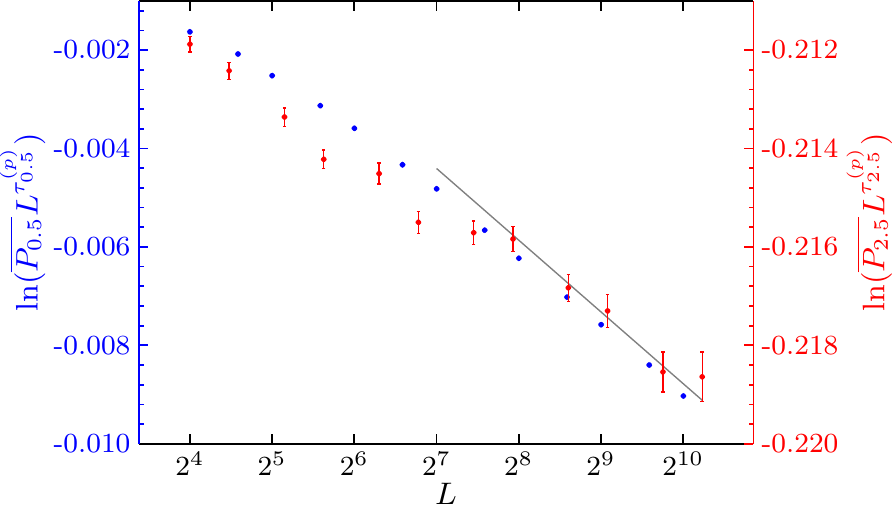}
	\caption{Average IPR reduced by parabolic scaling relation for the reciprocity pair $q=1/2$ and $q=5/2$ as function of the system size $L$.  The line is a guide to the eye with slope corresponding to a $\gamma$-value of 0.002244.
	\label{fig11}}
\end{figure}Notice that these  deviations as identified in Fig.  \ref{fig10} (inset, bottom panel) are consistent with reciprocity in the sense that $\text{redP}_{q}, q=1/2,5/2$ tend towards sharing the same slope; we highlight this important feature in Fig. \ref{fig11} displaying the reduced average IPR for $q=1/2,5/2$.
While finite-size corrections occur in both traces, the curves are seen to follow a common trend.

We remark that for $q{=}1/2$ the data in Fig. \ref{fig11} exhibits a clear  curvature to the right. In the case of parabolicity ($\gamma=0$), the opposite trend is expected, i.e. a curvature to the left indicating a flow towards a horizontal line. We take this as further evidence that non-parabolicity, though numerically small, is a robust feature of our data.  

\subsubsection{The case $q\approx 0$} 
\begin{figure}[t]
	\includegraphics{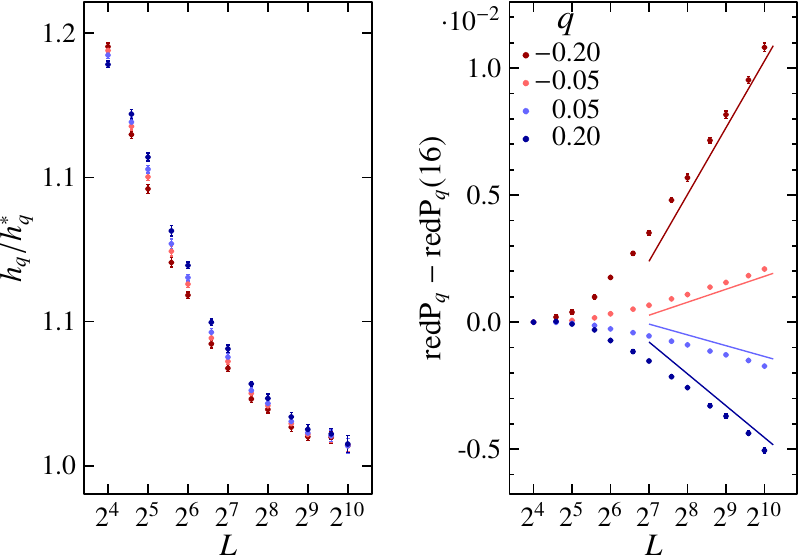}
	\caption{ Data analogous to Fig. \ref{fig2} for $q=\pm1/20,\pm1/4$
	Left panel: Normalized distribution height $h_q/h_q^\ast$ as function L. The asymptotic value $h_q^\ast$ is $19.32, 89.35, 29.85, 99.47$ for $q=\pm1/5, \pm1/20$, respectively. Right panel: Shift of the peak-position with system size $L$. The solid lines guide eye; the slopes correspond to a $\gamma$-value 0.002244. 
	\label{fig12}}
\end{figure}
We continue the analysis with a moment near zero, i.e. $q=\pm1/20$ and $q=\pm1/4$, at which according to \eqref{e12} corrections to parabolicity are expected to become even stronger than at $q=1/2$. Due to the increased statistical uncertainty of the symmetry partners, $q=59/20,11/4$, we here focus on the small $q$-regime.  
\begin{figure}[b]
	\includegraphics{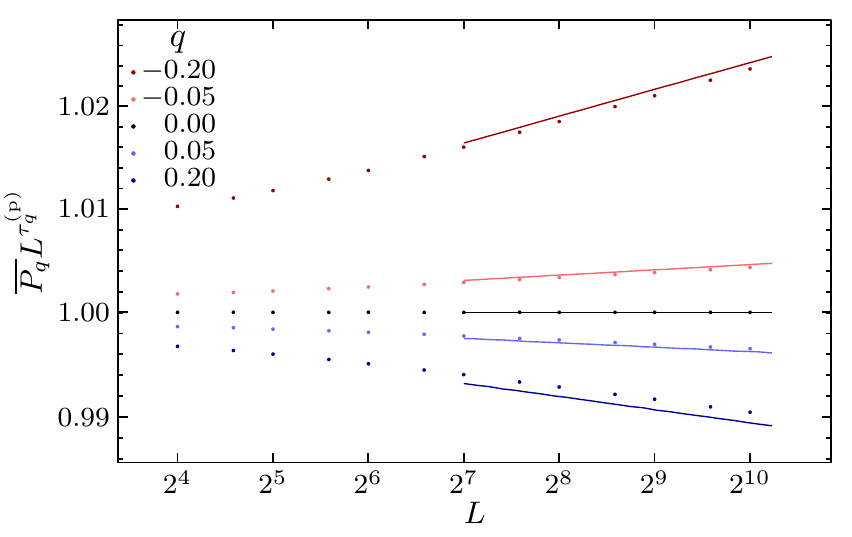}
	\caption{Plot similar to Fig. \ref{fig4} for $q=\pm1/20,\pm1/4$ and $q=\pm1/4$ in logarithmic scale for both axes. 
	Solid lines guide  the eye with slope corresponding to  a $\gamma$-value 0.002244. 
	\label{fig13}}
\end{figure}
The trend that has announced itself already at $q=1/2$ here consolidates: 
In Fig. \ref{fig12} the height $h_q(L)$ 
shows a fast convergence behavior that has a counterpart in the reduced position,  $\text{redP}_q(L)$, only if a residual flow - and hence deviations from parabolicity - are admitted. The scaling of the reduced average IPR, $\overline{P}_q(L)$ confirms this picture, see Fig. \ref{fig13}. 

\begin{figure}[t]
	\includegraphics{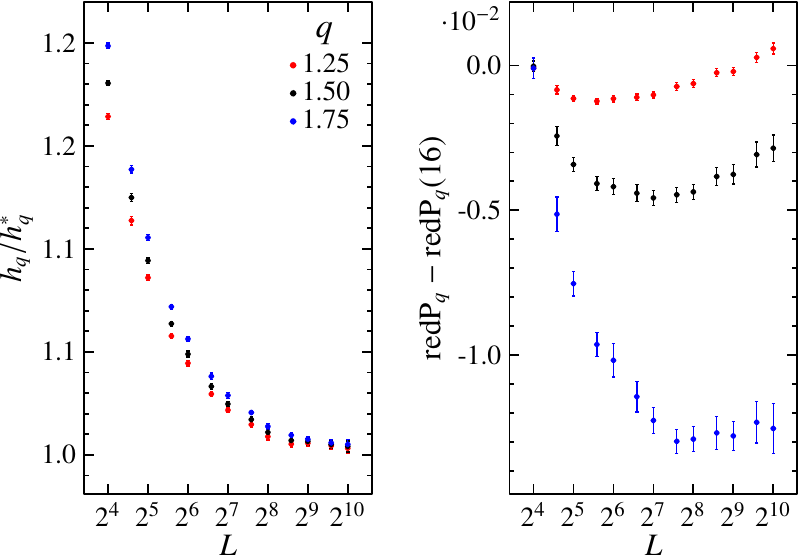}
\caption{	Normalized distribution height, $h_q/h_q^\ast$, (left panel) and relative shift of the peak position of the reduced distribution, $\text{redP}_q$, (right panel) as function of $L$ for  $q=5/4,3/2,7/4$. Error bars represent statistical uncertainties due to limited ensemble averaging and fluctuations based on fit ranges. The asymptotic value $h_q^\ast$ is $16.35$, $6.848$, $3.942$ for $q=5/4,3/2,7/4$, respectively.
	\label{fig15} }
\end{figure}
\subsubsection{The case of $q\in[1,2]$} 
As the final discussion of individual $q$-values we consider the symmetry point $q=3/2$ and the pair  $q=5/4,7/4$ around it as representatives of the region $q\in [1,2]$. The characteristic flow parameters are given in Fig. \ref{fig15}. 

The flow of $h_q(L)$ (and similar also for $\sigma_q(L)$, not shown)  exhibits the familiar convergence consistent with $y{\approx}1$. In contrast, the reduced shift behaves in a  non-monotonous way, which by itself does not suggest converged behavior.    
The situation becomes clear after consulting Fig. \ref{fig16}. 
\begin{figure}[b]
	\includegraphics{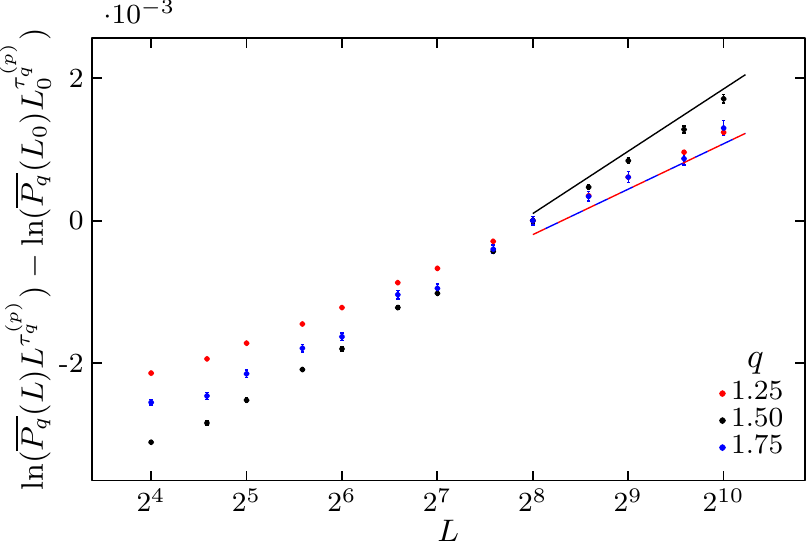}
	\caption{Average reduced IPR for $q=5/4,3/2,7/4$ in log-representation of both axes. The reference scale $L_0=2^8=256$ is the beginning of the asymptotic regime as indicated from the overlapping traces of the reciprocity pair, $q=5/4,/7/4$ and the convergence of the peak shape, see Fig. \ref{fig15}. The  asymptotic slope indicates deviations from parabolicity, i.e., $\Delta x_q{\coloneqq} x_q{-}x_q^\text{(p)}$ . The asymptotic collapse of $q=5/4,7/4$ reflects the symmetry with respect to $q=3/2$. Solid lines guide  the eye. 
	 \label{fig16}
	 }
\end{figure}
It shows that the traces for $q=5/4,7/4$ take the same slope at $L\gtrsim 300$, i.e., in the regime where $h_q(L)$ saturates. The non-vanishing slope is a manifestation of non-parabolicity. Also, it is seen that the trace corresponding to $q=3/2$ is intersecting with the other traces exhibiting a larger slope. From this trace we extract $\gamma_{3/2}= 0.00220\pm 00005$ consistent with the other estimates. 

\section{Asymptotics over the full spectral range} 
We extend the findings made for selected $q$-values over the entire range. 

\subsection{ IPR-scaling and hyper-collapse}
For a global description of the IPR, we define a function 
\begin{equation}
    F_q(L)\coloneqq 8 \ln(\overline{P}_q(L)\ L^{\tau_q^p})/[q(3-q)]. 
    \label{e14} 
\end{equation}
$F_q(L)$ is expected to scale as  
\begin{eqnarray}
F_q(L) &\approx& F_q(L_0) + \frac{x_q - x_q^\text{(p)}}{q(3-q)/8}\ln L/L_0, \nonumber\\
&=& F_q(L_0)+ 8\gamma (q-1)(q-2) \ln L/L_0. \nonumber
\end{eqnarray} 
where $L_0$ denotes a  reference length that indicates the beginning of the asymptotic scaling regime. The  numerical data corresponding to $F_q(L)$ is displayed in Fig. \ref{fig17} with $L_0=256$. 
At $L\approx L_0$ two traces that correspond to $q$-values paired via the reciprocity symmetry $x_q=x_{3-q}$ coalesce within the numerical error bars that represent statistical noise. 
By analyzing the scaling of the entire distribution function we have argued before that this is also the system size that indicates the onset of the asymptotic scaling regime. Therefore, we interpret the slope of $F_q(L)$ seen at $L>L_0$ in Fig. \ref{fig17}, upper panel as an evidence for the existence of non-parabolic corrections in $\tau_q$. 
\begin{figure}[b]
	\includegraphics{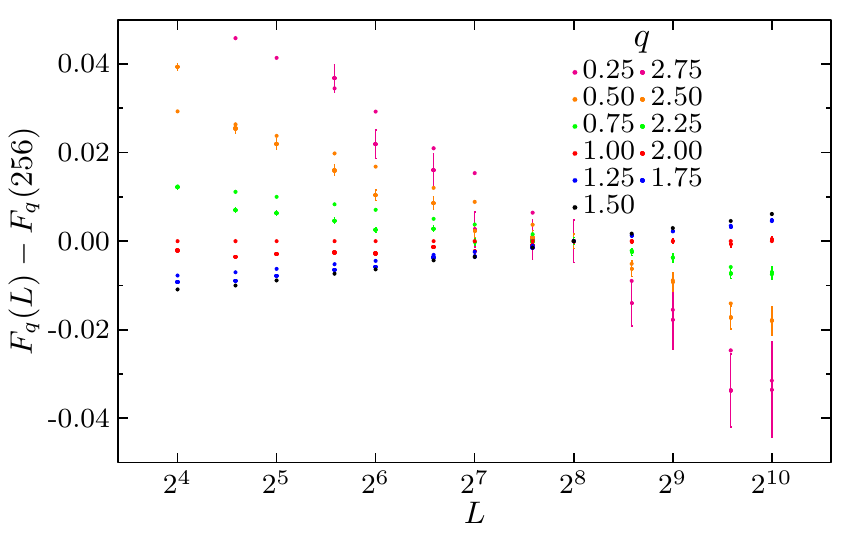}
	\caption{ The function $F_q$  defined in Eq. \eqref{e14}. At system sizes $L\gtrsim L_0, L_0=256$ it displays the expected collapse of reciprocity pairs. \label{fig17}}
\end{figure}

For quantitative estimates of non-parabolicity, we employ the scaling ansatz separately for each $q$ with fixed (universal) $y$   
\begin{equation}
    \overline{P}_q(L) L^{\tau_q^p} = L^{-\Delta\tau_q} \sum_{j=1}^{N_y} a^{(j)} L^{-jy}. 
    \label{e14a} 
\end{equation}
\begin{figure}[t]
	\includegraphics{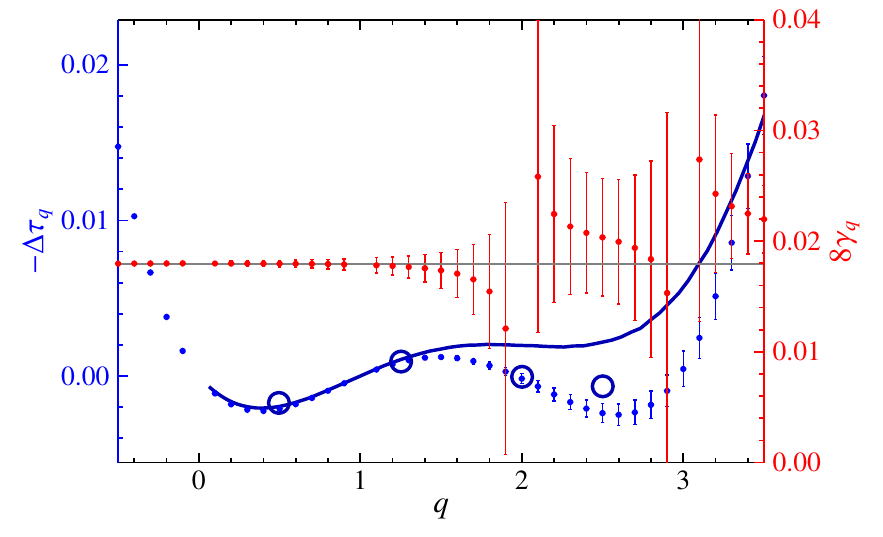}
	\caption{Estimate of $\Delta\tau_q$ and  corresponding $\gamma_q$ as a function of $q$. Fits are based on $\overline{P_q}L^{\tau_q^{(p)}}$ with $L{\geq}32$, expansion $N_y{=}2$, and fixed $y{=}0.75$. The horizontal line corresponds to $8\gamma=0.01795$. For comparison, earlier data by~\textcite{Evers2003} for $\Delta\tau_q$ is also shown (solid blue line and open symbols). 
	\label{fig18}}
\end{figure}
 \begin{figure}[b]
	\includegraphics{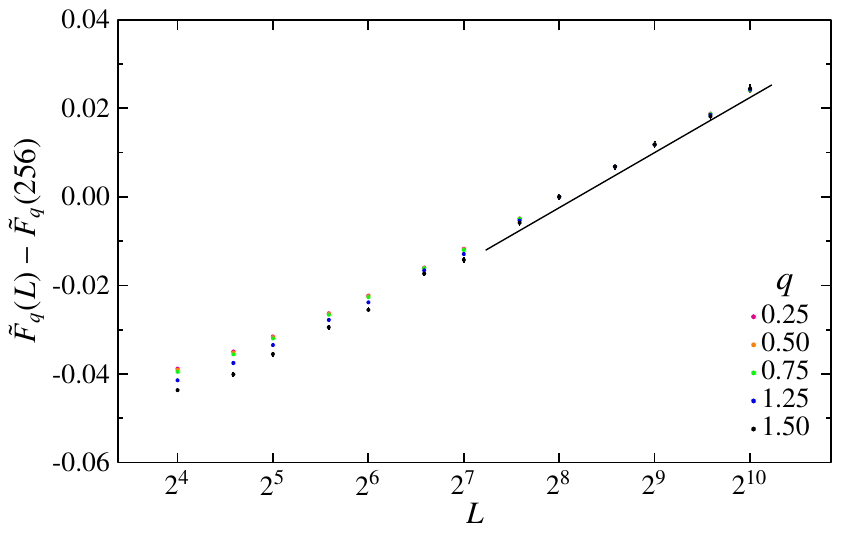}
    \caption{ The function $\tilde F_{q}(L)$ displays a nearly perfect scaling collapse for all $q$-values listed (``hyper-collapse"). The finite slope of this trace is a manifestation of quartic terms in the multifractal spectra. The hyper-collapse indicates that corrections to quartic terms are very small in the regime of $q$-values here considered. \label{FirstFig19}
}
\end{figure}
By varying the fit initial conditions, we ensure the convergence to a global minimum. With this approach, we created several sets by manually varying the fit range, $N_y$, and especially $y$. For the later we considered values between $0.1$ and $1.5$. For these data sets we compared the overall fit quality and checked the quality of the numerical agreement with the exact values for $q=2$ and $q=3$  ($\Delta\tau_2=\Delta\tau_3=0$). In particular $\Delta\tau_3=0$ provides a strong figure of merit to discriminate between values of $y$. Based on the current data, we observe reasonable fit parameters for $0.5\lesssim y\lesssim 1.0$. 
A set of fit parameters thus obtained for $\Delta \tau$ is displayed in Fig. \ref{fig18}. 
Based on the individual fit parameter $\Delta\tau_q$,  the quartic scaling factor $\gamma_q$ has been estimated; the result is also displayed in Fig. \ref{fig18}. 

A remarkable property of the function $\gamma_q$ thus obtained is its weak dependency on its argument, $\gamma_q=\gamma$, with $8\gamma=0.0178\pm 0.0002$; the error bars are discussed in App. \ref{aB}.
This observation motivates the definition of the scaling function
\begin{eqnarray}
\tilde F_q(L)&\coloneqq& F_q(L)/[(q-1)(q-2)]
\nonumber\\ &\approx& \tilde F_q(L_0) + \frac{x_q - x_q^\text{(p)}}{q(q-1)(q-2)(3-q)/8}\ln L/L_0 \nonumber\\
&=& \tilde F_q(L_0)+ 8\gamma \ln L/L_0 \nonumber 
\end{eqnarray} 
plotted in
Fig.~\ref{FirstFig19}.
It displays the striking feature of all reciprocity pairs collapsing onto the same master curve - within the numerical error bars (``hyper-collapse''). 

\subsection{Width $\sigma_q(L)$ and height $h_q(L)$}

As natural descriptor of the form of the distribution function $\mathcal{P}_q(\ln P_q;L)$ we have employed the second moment $\sigma_{q}(L)$ and the peak height $h_q(L)$. 
Following Eq. \eqref{e8} we analyze the finite-size corrections for each $q$, thus estimating the fixed point values 
$\sigma_q^{*}, h_q^{*}$. 
\begin{figure}[t]
	\includegraphics{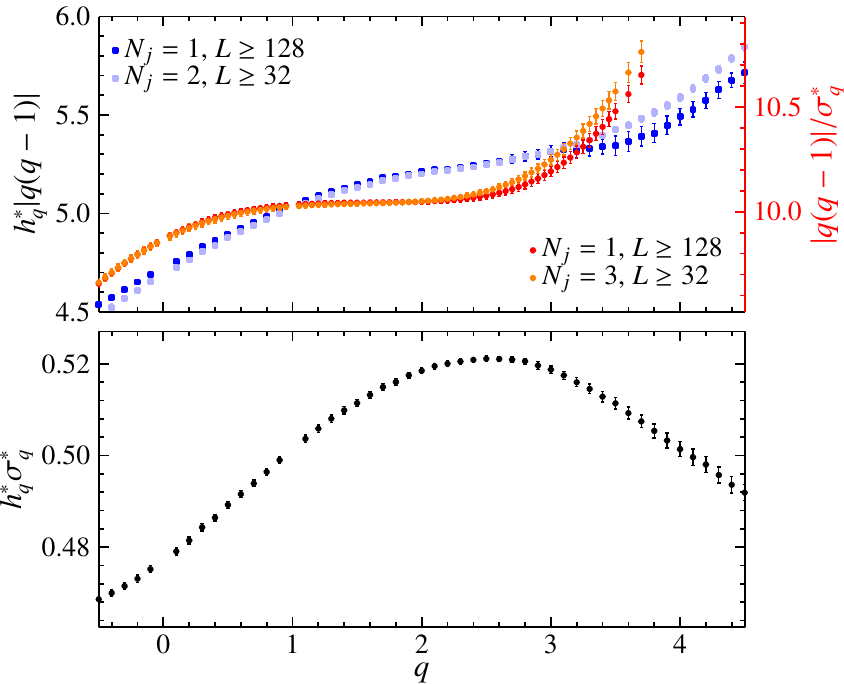}
	\caption{Distribution height $h^*_q$ and inverse width $1/\sigma^*_q$ as function of $q$. The scaling by $|q(q-1)|$ removes the trivial divergences and zeros at $q=0,1$. The two data sets are shown per observable in order to illustrate the goodness of the fitting, see Appendix \ref{aB}. The lower panel shows the effective area $h^*_q\sigma^*_q$.
	\label{fig18afocus}}
\end{figure}
Figure\;\ref{fig18afocus} shows the height $h_q$ and inverse width $1/\sigma_q$ as function of the moment $q$ reduced by the behavior  near $q=0,1$; by definition: $1/h^*_q=0, \sigma^*_q=0$.  
As readily seen from the data, the product $h_q^*\sigma^*_q$ depends on $q$. It thus is indicating a gradual change of the asymptotic shape of $\mathcal{P}_q$ with varying $q$.

\subsection{Tail exponents\label{ss:tail}} 
The evolution of the IPR distribution with $q$ also manifests its 
\begin{figure}[t]
	\includegraphics{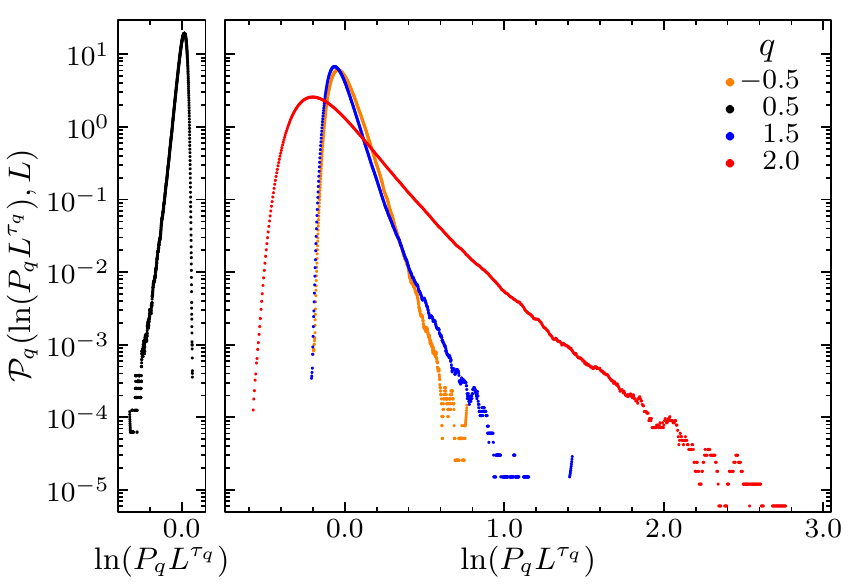}
	\caption{Distribution ${\mathcal P}_{q}(\ln\left[P_{q}(L)L^{\tau_{q}}\right];L)$ at $L=1024$ for representative $q$ values. The tail is seen to be approached from the peak value passing through an inflection point. 
	Fig. \ref{fig20} zooms into the tail. }
	\label{fig19}
\end{figure}
asymptotic regime where it is described by a power-law 
\cite{Mirlin2000,Evers2008RMP} 
\begin{equation}
    \mathcal{\tilde{P}}_q(P_q;L)\sim \mathcal{P}_q(\ln P_q;L)/P_q \sim P_q^{-1-\rmx_q}.
    \label{e15} 
\end{equation}
with tail exponent $\rmx_q$, see Figs. \ref{fig3} and 
\ref{fig19} for illustration. 
The asymptotic regime is given  with 
$P_q/P_q^\text{typ}\ll 1, q\in(0,1)$ and 
$P_q/P_q^\text{typ}\gg 1$ otherwise. 

{\bf Typical and average IPR.} 
For $\rmx_q>1$, the first moment of the distribution exists, and hence the average $\overline{P_q}$ and the typical value $P^\text{typ}_q$ show the same scaling with the system size. Contrary, at $\rmx_q<1$ the moments will be dominated by the upper bound of the integral, which depends on $L$, so average and typical IPR will scale differently \cite{Mirlin2000,Evers2008RMP}. 
At the critical point
$\rmx_{q^\pm}=1$, which separates  both regimes the Legendre-transformed $\tau_q$ vanishes, $f_{q^\pm}=0$, where 
\begin{eqnarray}
    f_q &\coloneqq& q\frac{\partial \tau_q}{\partial q} -\tau_q = d +q\frac{\partial x_q}{\partial q} - x_q \nonumber\\
    &=& d -\frac{1}{8} q^2 - \gamma q^2 (3q^2-12q +11) + \mathcal{O}(q^5). 
    \label{e16}
\end{eqnarray}
For $\gamma \geq 0$, the polynomial has two real roots, $\pm4$ in the case $\gamma=0$. For the realistic value $\gamma=1/448$, we obtain $q_-^*{=}-2.714$ and $q_+^*{=}3.739$. 
We mention in passing that the freezing limit is given by
$\partial \tau_q/\partial q=0$; it describes the upper bound in $q$ for the validity of Eq. \eqref{e16} \cite{Mirlin2000,Evers2008RMP}. We estimate $q_c\approx 5.90$ for $8\gamma=0.01786$, so the evolution at our considerations are safely away from this limit.
\begin{figure}[t]
	\includegraphics{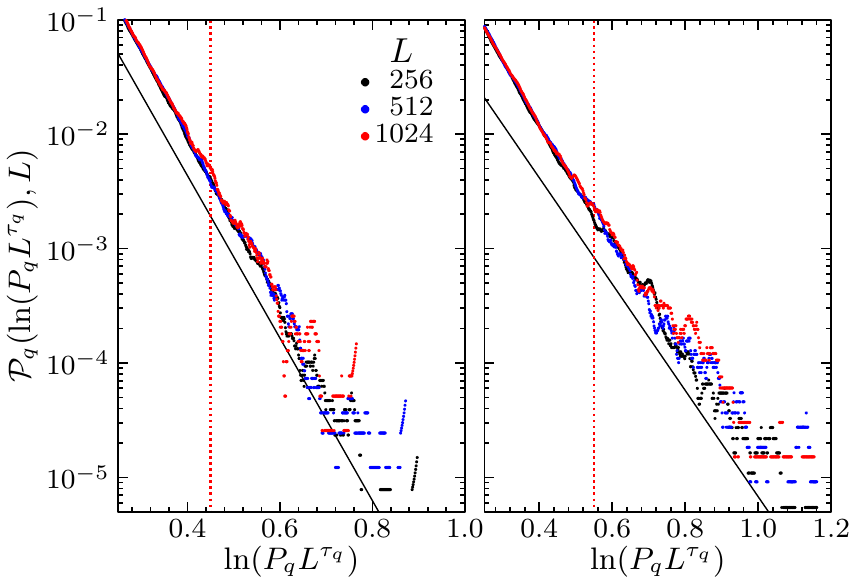}
	\caption{Evolution of the tails of the distribution ${\mathcal P}_{q}(\ln\left[P_{q}(L)L^{\tau_{q}}\right];L)$ with increasing system size $L{=}256, 512, 1024$ at $q=-1/2,3/2$; raw data shown in Fig. \ref{fig19} . The solid lines indicate fitted power laws with $\rmx_{-0.5}=10.7(4)$ and $\rmx_{1.5}=16.4(4)$; 
	the red dotted line indicates the fitting window.}
	\label{fig20}
\end{figure}
\begin{figure}[b]
	\includegraphics{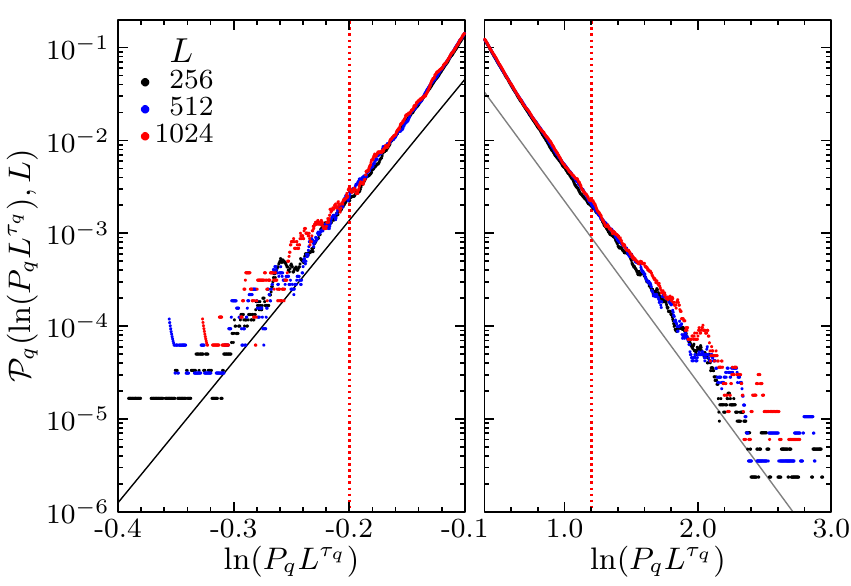}
	\caption{Similar to Fig.\;\ref{fig20}, but for $q=1/2$ (left) and $q=2$ (right) with $\rmx_{0.5}=-35(1)$ and $\rmx_{2}=4.5(1)$.}
	\label{fig21}
\end{figure}

{\bf Numerical estimates.} 
Figure \ref{fig19} shows the tail of the distribution $\mathcal{P}_q$ and the corresponding inflection point for selected $q$-values. A zoom-in on the corresponding tails is displayed in Figs. \ref{fig20} and \ref{fig21}. The data shown in these figures highlights the difficulties encountered when trying to numerically estimate the tail exponent $\zeta_q$: (i) The tail exhibits a slow evolution with increasing system size $L$ tending towards decreasing slope. (ii) The power law is best developed far in the tails, where rare events prevail and statistical noise is large. 
When fitting the tail exponents, we have restricted the fitting window to the regime in $\mathcal{P}_q$ outside the inflection point. Because of (ii) the numerical error bars are sizable, because of (i) our estimate should be considered an upper bound, strictly speaking. 
The results for the tail exponents obtained in this way are given in Fig. \ref{fig22}. 

{\bf Discussion.} As seen already from the raw data, Fig. \ref{fig19}, $\zeta_q$ is a rapidly increasing function when approaching $q=1$ from above. Moreover, it displays a change in sign at $q=1$ and, similarly, also at $q=0$. These observations have motivated us to plot in Fig. \ref{fig22} the product $\zeta_q q(q-1)$, which is always positive and appears to display a weaker dependency on $q$ - at least for $q$-values sufficiently far away from $q=0,1$. Near these particular values, $\zeta_q$ becomes very large and hence the numerical estimates carry very large error bars. 
At $q>1$ the exponents $\zeta_q$ display the same qualitative behavior already known from other Anderson transitions \cite{Evers2008RMP}: $\zeta_q$ is decreasing with increasing $q$ for $q>1$. 

\begin{figure}[]
	\includegraphics{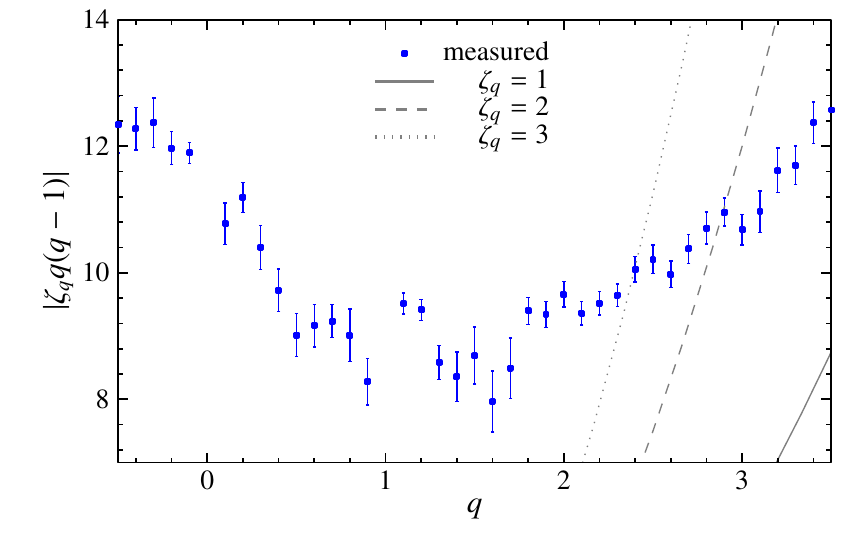}
	\caption{Tail exponents $\rmx_q$ of the IPR-distribution as defined in Eq. \eqref{e15}. Three lines are also shown. There intersection with the data trace indicates the $q$-value at which the tail exponent takes values $\rmx_q=1$,$2$,$3$. At $q$-values larger than these the  $1$\textsuperscript{th}, $2$\textsuperscript{nd}, $3$\textsuperscript{rd} moment of the IPR will be dominated by the integral boundaries rather than by the bulk of the distribution.  
	}
	\label{fig22}
\end{figure}

\section{The collapse of finite-size distorted distribution functions}

The shape of the distribution function $\mathcal{P}_q(\ln P_q(L)L^{\tau_q};L)$
exhibits sizable finite-size effects seen, for instance, in Figs. \ref{fig1}, \ref{fig5} and \ref{fig8}. In this section we present a heuristic single-parameter rescaling of this distribution to the effect that all traces seen, e.g., in Fig. \ref{fig1} collapse onto a single master curve. 

\subsection{Single-parameter rescaling of distribution functions} 
We consider the integrated distribution function  
\begin{equation}
    {\mathcal N}_q(\ln\left[P_q(L)L^{\tau_q}\right];L) \coloneqq \int_{-\infty}^{\ln P_q L^{\tau_q}}\mathrm{d}x\, \mathcal{P}_q(x;L)\quad, 
    \label{e18b}
\end{equation}
so ${\mathcal P}_q(x;L) = \partial_x{\mathcal N}_q(x;L)$. 
The species corresponding to $q{=}2$ is depicted in Fig. \ref{fig23} and $q=1/2$ in Fig. \ref{fig23b}. The data is seen to exhibit a common crossing point, e.g.,   
$\ln P_2L^{\tau_2} \approx -0.042 \pm 0.004$ in Fig. \ref{fig23}. Clearly, the existence of a crossing point of a pair of traces corresponding to two neighboring system sizes is expected. Also expected is a flow of the crossing point with increasing system sizes towards a limiting value.  Therefore it is remarkable that there is virtually no such flow discernible in the inset of Fig. \ref{fig23} even though the raw data, Fig. \ref{fig1}, does exhibit sizable finite-size effects of the order of 20\%. 

The stability of the crossing point allows for an attempt at a single-parameter rescaling of the abscissa in Fig. \ref{fig23} with the crossing point, $c_q$, being the fixed reference position:
\begin{equation} 
{\mathcal N}_q(\Lambda_q(L);L)=
{\mathcal N}^\infty_q(\lambda^{-1}_q(L)\{\Lambda_q(L)-c_q\}+c_q),
\label{e18} 
\end{equation} 
where $\Lambda_q(L)\coloneqq\ln\left[P_q(L)L^{\tau_2}\right]$.
A natural choice for the scale factor, $\lambda_q(L)$, here introduced would be the (inverse) slope at the crossing point. We note that in Fig. \ref{fig23} the crossing point turns out to be very close to the inflection point, where the slope is given by the height $h_q(L)$. With  this observation, we adopt the definition  $\lambda_q(L){\coloneqq} h_q^*/h_q(L)$ constructed so that $\lambda_q^*{=}1$. 

As is demonstrated in Fig. \ref{fig23}, left panel, the rescaled integrated  distribution function, Eq. \eqref{e18}, for $q{=}2$ exhibits a nearly perfect collapse towards a master curve in a window of system sizes, $L$, that covers almost two decades. This is highly remarkable, because apart from reading out $h_q(L)$ in Fig. \ref{fig1} there is no fitting parameter involved. Only at larger arguments deviations from the master curve are visible for the smallest system sizes.
\begin{figure}[t]
	\includegraphics{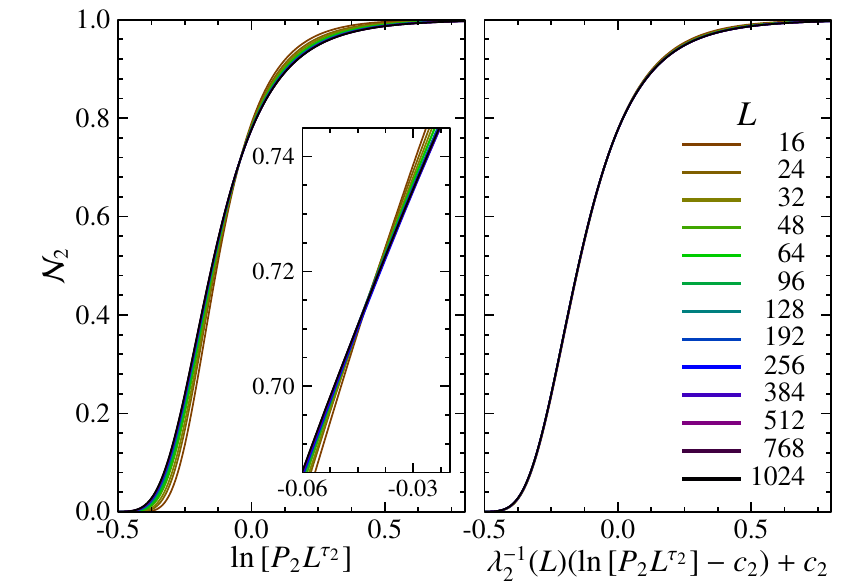}
	\caption{Integrated distribution ${\mathcal N}_2(\ln\left[P_2(L)L^{\tau_2}\right];L)$ before (left) and after (right) abscissa rescaling about the common intersection point $c_2{=}-0.042 \pm 0.004$. Inset: Vicinity of the crossing point. The scale parameter abbreviates $\lambda_2(L)\coloneqq h_2^*/h_2(L)$ with $h_2(L)$ given in Fig.\;\ref{fig2}.
	\label{fig23}}
\end{figure}

A collapse of similar quality can be obtained also at other $q$-values, e.g., for $q{=}1/2$ as demonstrated in Fig. \ref{fig23b}. As also shown for this case, a collapse can only be achieved if quartic terms in $\tau_q$ are accounted for: when stipulating $\gamma{=}0$ traces corresponding to different system sizes do not exhibit the crossing point (inset Fig. \ref{fig23b}).

{\bf Further discussion.} To further investigate the heuristic rescaling we here propose, we plot in Fig. \ref{fig24} the distribution functions shown in Fig. \ref{fig3} after rescaling, which correspond to the derivative of the traces shown in Fig. \ref{fig23}, right: while the collapse in the bulk of the distribution function is close to perfect, deviations in the tail can be seen also, here. 
\begin{figure}[t]
	\includegraphics{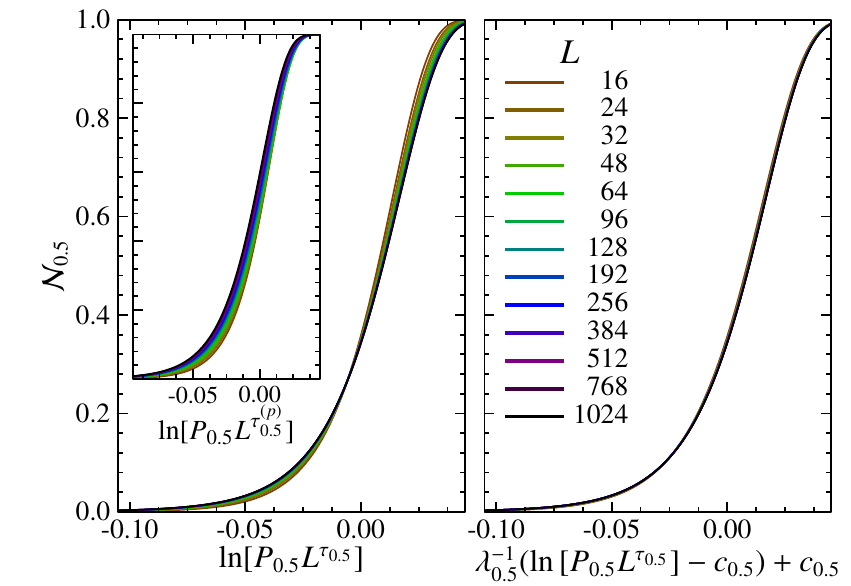}
	\caption{Similar to Fig. \ref{fig23} but for $q=1/2$ with the common intersection point $c_{0.5}{=}-0.005\pm 0.002$. Inset: Integrated distributions as they would have been obtained with $\gamma=0$, i.e. ${\mathcal N}_{0.5}(\ln\left[P_{0.5}(L)L^{\tau^{(p)}_{0.5}}\right];L)$. The plot highlights the importance to account for quartic corrections to $\tau_q$ to achieve the data collapse seen in the right-hand side panel.
	\label{fig23b}}
\end{figure}

The single-parameter ansatz \eqref{e18} implies for the distribution function Eq. \eqref{e6a}, i.e. 
\begin{equation} 
{\mathcal P}_q(\Lambda_q;L)=\frac{\lambda^*_q}{\lambda_q}\
{\mathcal P}^\infty_q\left(\frac{\lambda_q^*}{\lambda_q}\{\Lambda_q-c_q\}+c_q\right),
\nonumber
\end{equation} 
where the dependency of $\Lambda_q(L)$ and $\lambda_q(L)$ on $L$ has been suppressed in our notation.  
The expression allows for an interpretation of finite-size corrections as far as they affect the bulk of the distribution - rather than its tail; they manifest as a ``dressing" of the reduced IPR amplitudes
\begin{equation}
P_qL^{\tau_q} e^{-c_q} \rightarrow [P_qL^{\tau_q}e^{-c_q}]^{\lambda_q^*/\lambda_q}.
\end{equation}
For the average amplitude we thus derive 
\begin{equation}
    \overline{P_q L^{\tau_q}}(L) = e^{(\lambda^*_q/\lambda_q-1)c_q}\int dx \ e^{\lambda_q x}\ \mathcal{P}_q^\infty(x).  
    \label{e21} 
\end{equation}
Upon expanding the right-hand side of Eq. \eqref{e21} in $\lambda_q(L)-\lambda_q^*$ we recover the form Eq. \eqref{e14}. Judging from the excellent collapse achieved in Figs. \ref{fig23} and \ref{fig24}, the expression appears to have the advantage that it partially resums the higher-order terms in \eqref{e14}. 

\subsection{Exponent fittings via Kolmogorov-Smirnov test}

The preceding analysis of the flow of distribution functions motivates a fresh approach towards estimating multifractal spectra, $\tau_q$, in the  presence of strong  finite-size corrections. Based on field-theoretic arguments \cite{Wegner1976,CardyBook}, the conventional method follows Eq. \eqref{e14} fitting the average IPR, $\overline{P}_q(L)$, with a leading power and subleading corrections.\cite{Evers2008RMP} From a computational perspective a frequent problem with this procedure is that fits are unstable due to a proliferating number of fitting parameters. 
\begin{figure}[t]
	\includegraphics{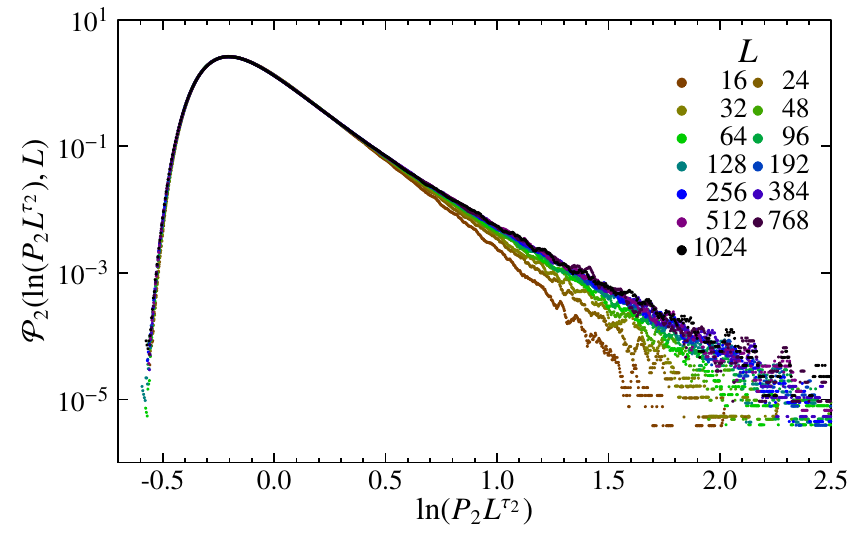}
	\caption{Distribution function $\mathcal{P}_q$ shown in Fig. \ref{fig3} here after performing the single-parameter scaling transformation. The traces correspond to the  derivative 
	of the integrated distribution $\mathcal{N}_q$ shown in Fig. \ref{fig23} [see Eq. \ref{e18b}]. The plot highlights the excellent collapse seen in the bulk of the distribution and the lack thereof in the tails. }
	\label{fig24}
\end{figure}

{\bf Method.} We here propose an alternative method for estimating exponents $\tau_q$. It is simple and as compared to the conventional approach it has the dramatic advantage that except for $\tau_q$ there is no other parameter that requires optimization. The main idea is to define a family of auxiliary functions
\begin{equation}
    \mathcal{N}_q(\ln\left[P_q(L)L^{\tau}\right];L),
    \nonumber
\end{equation}
with family parameter $L$; examples for two families that differ by the choice of $\tau$ have been depicted in Fig. \ref{fig23b}. With increasing $L$ family numbers become more and more indistinguishable, if and only if $\tau$ coincides with $\tau_q$. To monitor this evolution we define a distance between two family members: 
\begin{equation}
    \mathfrak{D}_q(L,L';\tau) = \text{max}[  \mathcal{N}_q(\Lambda_q^\tau(L);L)  - \mathcal{N}_q(\Lambda_q^\tau(L^\prime);L^\prime)
    ] \label{e22} 
\end{equation}
where we have abbreviated $\Lambda_q^\tau(L){\coloneqq}\ln\left[P_q(L)L^{\tau}\right]$.  We employ this particular measure of closeness because it allows us to adopt the Kolmogorov-Smirnov test 
\cite{DeGroot2012} to assess the statistical significance, traditionally called 
\begin{figure}[t]
	\includegraphics{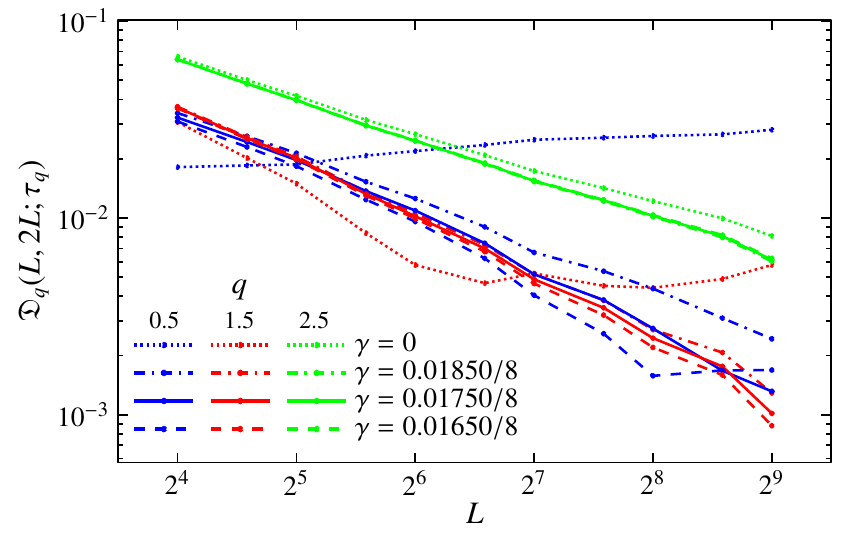}
	\includegraphics{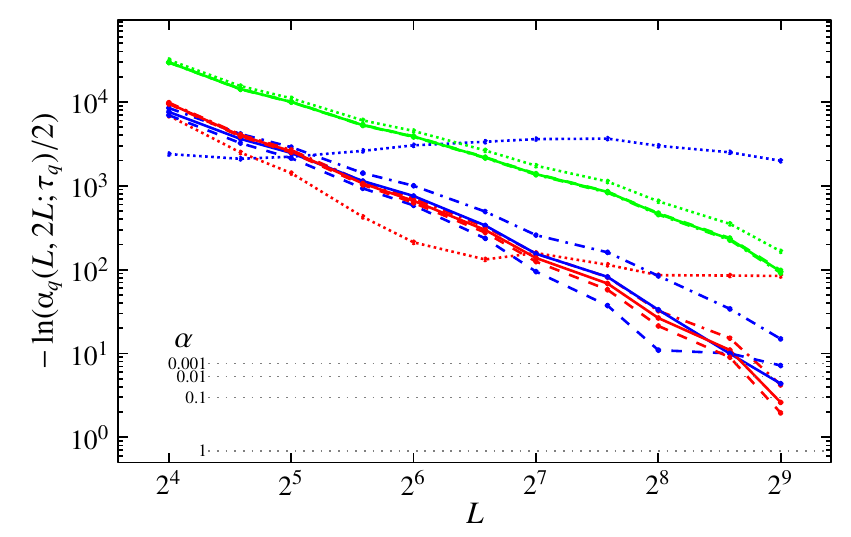}
	\caption{Estimating the spectrum $\tau_q$ adopting the Kolmogrov-Smirnov test. Upper panel: Distance for a pair of IPR-distributions,  $\mathfrak{D}_q(L,2L;\tau_q)$, defined in Eq. \eqref{e22} at system sizes $L$ and $2L$ for different $\tau_q$ ``guesses" with and without quartic terms:  $\gamma=0$ (dashed lines) and $8\gamma=0.01750$ (solid). The ``best-estimates" for $\tau_q$ are those with smallest distances at largest system sizes. Lower panel: Significance level $\alphaup_q(L)$ as given in Eq. \eqref{e23} for every distance given in the upper panel. The horizontal line indicates significance levels of 0.1\%, 1\% and 10\% ($\alphaup=10^{-3},10^{-2},10^{-1}$). 
	\label{fig26}}
\end{figure}
$\alphaup$, of a distance obtained for a pair of two numerical data sets: 
\be
\label{e23} 
\alphaup_q(L,L',\tau)=2\exp\left(- \frac{2 \Nsamples(L) \Nsamples(L')}{\Nsamples(L)+\Nsamples(L')} \mathfrak{D}^2_q(L,L';\tau) \right),
\ee
where $\Nsamples$ denotes the number of disorder configurations in the ensemble, see Tab. \ref{t2}. 
The ``best guess" for $\tau_q$ is given by the parameter $\tau$ that minimizes the distance between two neighboring pairs $L,L^\prime$ in the large-$L$ limit taken at $L^\prime/L$ fixed. A data point for the distance will be accepted if the corresponding significance is better than a predefined level, e.g.,1\%: $\alphaup\lesssim 0.01$.
As an illustration, Fig. \ref{fig26} displays the typical  evolution of $\mathfrak{D}_q(L,2L;\tau)$ with increasing system size. Here a quartic term  manifests as the superior choice as compared to a parabolic spectrum, $\gamma=0$. 

{\bf Discussion.}
The proposed approach to estimating $\tau_q$ operates by choosing a guess that brings the system-size flow of $\mathcal{P}_q(\ln P_q L^\tau)$ to a standstill in the limit of large $L$. The advantage of this approach is that the goodness of the guess can be read of from Fig. \ref{fig26} without fitting: worse guesses reveal themselves as compared to the better ones by leveling off to saturated values of the pair distance $\mathfrak{D}(L,L';\tau)$. Using this technique we arrive at an estimate $8\gamma_{1/2}{=}0.0175{\pm}0.0010$. 

\section{Summary and Outlook} 
The main goal of this work was to present an in-depth analysis of multifractality and finite-size corrections for the class-C quantum Hall transition that could serve as a paradigm for similar investigations in other symmetry classes. The symmetry class C lends itself most naturally for this purpose, because out of the full spectrum $\tau_q$ two 
nontrivial exponents, i.e. $q=2,3$, are known analytically and therefore can serve as a reference point for studying finite-size effects. 

As compared to most earlier studies, our investigation has not focused on average inverse participation ratios, $\overline{P}_q(L)$, but rather on the flow with system size $L$ of the entire distribution function $\mathcal{P}_q(\ln P_q;L)$. As it turns out, the shape of this function as far as its bulk is concerned is conveniently parametrized by a single parameter, e.g., its peak-value $h_q(L)$. We observe that the distribution 
$\mathcal{P}_q$ exhibits a scaling form that accounts for finite-size corrections with $\tau_q$ and $h_q(L)$ as the only input parameters.  

Embarking on this result, we have explored the potential of a novel approach to finite-size corrections based on the Kolmogorov-Smirnov test.
The method is sufficiently sensitive to allow us extracting $\tau_q$ essentially without any fitting to an accuracy good enough in order to reliably detect non-parabolic components in $\tau_q$: $\Delta\tau_q\coloneqq\gamma_qq(q-1)(q-2)(q-3)$. We obtain that $\gamma_q$ is essentially independent of $q$ with $8\gamma=0.0178\pm0.0010$ for $q\in(-0.5,3)$. This outcome satisfies the reciprocity symmetry.

We have confirmed these results by performing a standard analysis of finite-size corrections based on fitting  $\overline{P}_q(L)$ to a leading power law and irrelevant corrections: $\sim L^{-\tau_q}(1+\mathcal{O}(L^{-y}))$; irrelevant scaling indices could be reliably determined, $y\approx 0.9\pm0.3$ within a window  $q\in(-0.5,2)$; the large-$q$ bound is imposed by the loss of numerical stability at $q\gtrsim 2$. The origin of this loss has been traced back to the tail of the IPR distribution function; it is  characterized by an exponent $\zeta_q$ which falls below two, $\zeta_q<2$, at $q>\qplus$, so that the second moment of the IPR distribution is dominated by integral boundaries. 
The overall analysis fully confirms that terms of higher order than quartic are strongly suppressed in $\tau_q$. 

The versatile analysis techniques presented in this work are designed to readily carry over to other critical points. As an outlook, we mention that the quantum Hall transitions in symmetry classes A \cite{Zirnbauer2017,Zirnbauer2019} and AIII \cite{Sbierski2020} experience a resurge of attention, recently. 
It will be highly interesting to compare the critical behavior of these transitions that has been addressed previously by Evers et al.~[\onlinecite{Evers2008}] and Obuse et al.~[\onlinecite{Obuse2012}]  in greater depth, e.g., with respect to finite size corrections on distribution functions and with akin eye on the identification of the critical field theory.
Further, the Kolmogorov-Smirnov test advocated in this work as a methodological development is not without alternative in mathematical statistics. We here have to leave it to future work to unravel the full potential of this analysis method in the context of scaling and critical behavior near Anderson and quantum Hall transitions. 

\begin{acknowledgments}
 We thank Matthew Foster, Ilya Gruzberg, and  Alexander Mirlin for many discussions and useful comments on the manuscript. We also thank Ilya Gruzberg and Alexander Mirlin for earlier collaboration on closely related projects. 
F. E., D. H.-.P., and M. P. acknowledge support from the German Research Foundation (DFG) through the Collaborative Research Center, Project ID 314695032 SFB 1277 (project A03) and through the DFG project EV30/14-2.
SB acknowledges support from Department of Science and Technology~(DST), India, through Ramanujan Fellowship Grant No. SB/S2/RJN-128/2016, Early Career Award No. ECR/2018/000876, Matrics No. MTR/2019/000566, and MPG for funding through the Max Planck Partner Group at IITB. In particular, we acknowledge support of MPI-PKS, Dresden computing cluster support, where a part of the calculation is performed.
\end{acknowledgments}


%

\section{Appendix}
\subsection{Sample statistics\label{aA}} 
Table \ref{t2} lists the number of samples,  $\Nsamples$, for all (linear) system sizes $L$ considered.
 A number of $N_\text{ev}=6$ eigenvectors have been calculated per sample with eigenvalues taken closest to unity.  In a separate set of test calculations with $N_\text{ev}{=}4,6,12$ 
 we have ascertained that our results are not sensitive to the choice of $N_\text{ev}$;
 for a given disorder realization, the results for the IPR agree to an accuracy better than $11$ relevant digits at $L=128$. 
 For the statistics shown in the main paper only the eigenvector with eigenvalue closest to unity has been considered. As we show in  \ref{aC}, the other eigenvectors exhibit significantly larger finite-size corrections and therefore have been discarded from the main analysis. 
\begin{table}[h]
       \begin{tabular}{l|rrrrrrr}
    $L$ & 16 & 24 & 32 & 48 & 64 & 96 & 128	\\ 	\hline
    $\Nsamples(L)$ & 9216 & 6136 & 6000 & 6144 & 6750 & 6000 & 6000\\[2mm]
     $L$ & 192 & 256 & 384 & 512 & 768 & 1024 &\\ \hline
    $\Nsamples(L)$ & 6144 & 5582 & 5121 & 3632 & 2710 & 1930 &
    \end{tabular}
     \caption{Number of lattice realizations $\Nsamples$ (in units of $1000$) as function of system size $L$. \label{t2}}
\end{table}

\subsection{Details on the finite-size scaling analysis\label{aB}} 
\subsubsection{Irrelevant scaling corrections -- estimating $y$}
Corrections to scaling of the observables $\sigma_q(L)$ and $h_q(L)$ have been analyzed based on the expansion Eq. \eqref{e8}, i.e., 
\be
\sigma_q=\sigma_q^\ast\left(1+\sum_{j=1}^{N_y} \sigma^{(j)}_q L^{-jy}\ \right), \nonumber
\ee 
where $N_y$ denotes the expansion order. For assessing uncertainties in fitting parameters related to statistical and systematic errors, different combinations of regimes in $L$ and $N_y\leq3$ have been considered, as well as fits with $y$ kept adjustable or fixed. Both observables, $\sigma_q$ and $h_q$,  show a similar behavior with respect to the irrelevant exponent $y$ as well. We here focus on the $\sigma_q$-based data.  
\begin{figure}[t]
	\includegraphics{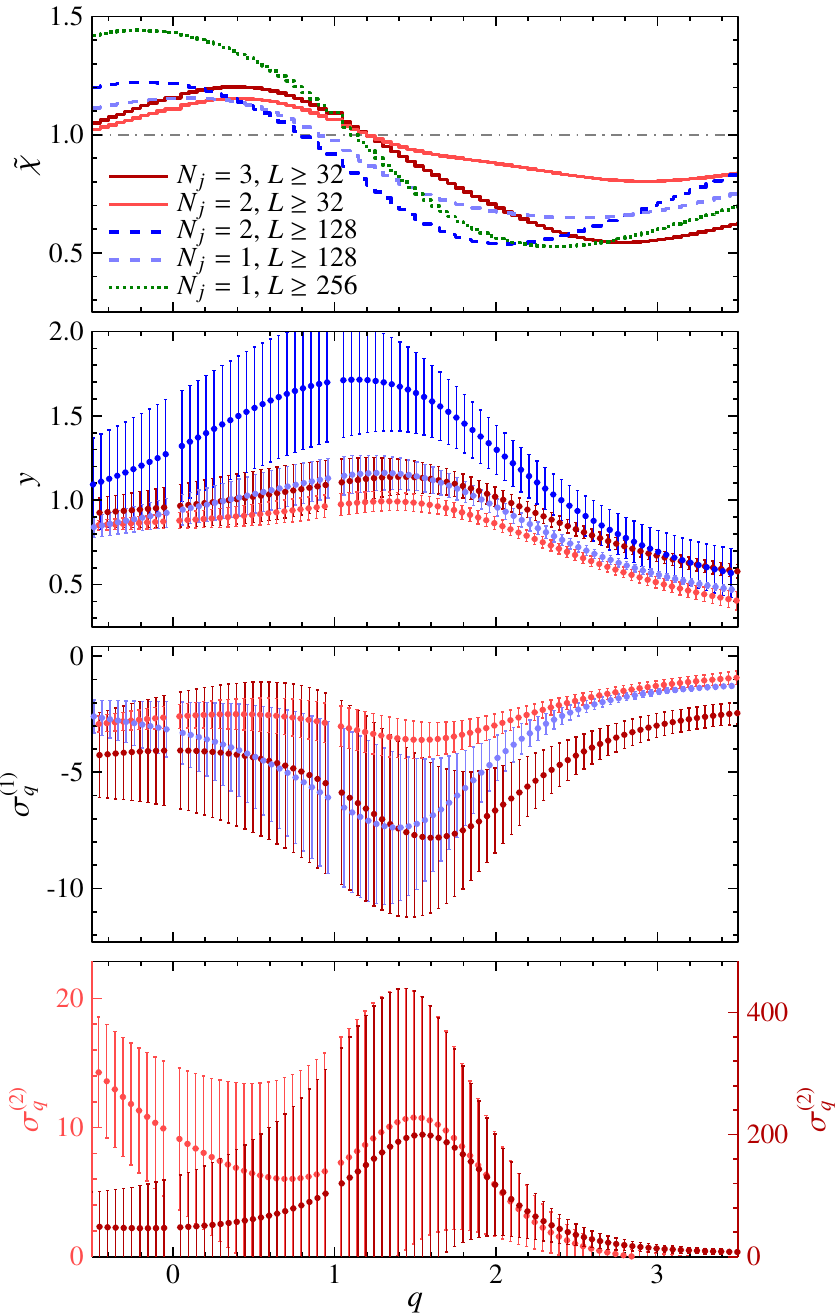}
	\caption{ Parameters following Eq. \eqref{e10} from fitting to $\sigma_q(L)$ data analogous to Figs. \ref{fig2} and \ref{fig6}. Estimates of the goodness of fit, $\tilde \chi_q$ (top panel),  irrelevant exponent $y$ (second panel) and amplitudes $\sigma_q^{(1,2)}$ (lower panel) are given for several fits with expansion order $N_y$ and system size $L$.
	}
	\label{fig27}
\end{figure}

In Fig. \ref{fig27}, we display the fitting parameters - goodness of the fit $\tilde\chi$, exponent 
$y$ and two amplitudes $\sigma^{(1,2)}$, - for various fitting conditions.  The estimates of different fits agree well with the error bars; the goodness of fit suggests that best results are obtained for $(N_y=2;L=32)$ and $(1;128)$. 
The results of Fig. \ref{fig27} are consistent with \eqref{e10} for $q\lesssim \qplus, \qplus\approx 2.7$ in the sense that in this regime the fit for $y$ is nearly the same for each moment $q$. At $q\gtrsim \qplus$, fits deviate from this expectation. Finite-size effects proliferate, which reflects in the fitting as estimates for $y$ reducing by a factor of two; 
the respective amplitudes $\sigma_q^{(1,2)}$ keep moderate values, see Fig. \ref{fig27}. 

\begin{figure}[]
	\includegraphics{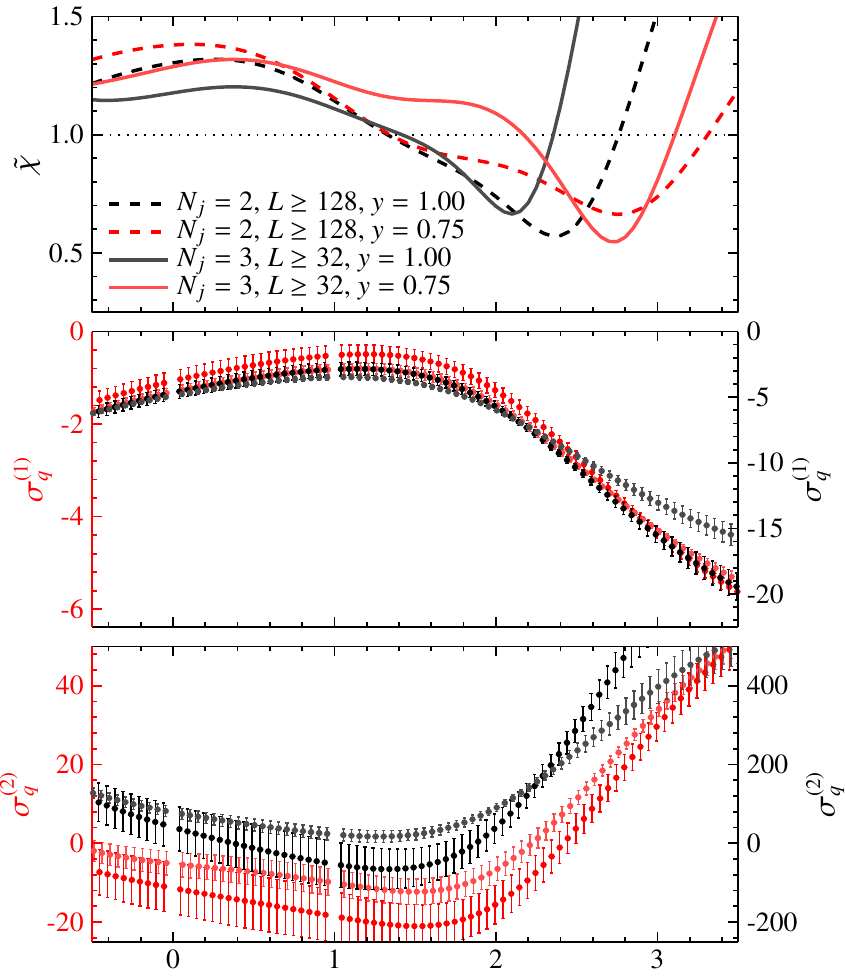}
	\caption{Fit quality $\tilde{\chi}$ as function of the moment $q$ for a $q$-independent irrelevant exponent $y$ for several fits with expansion order $N_y$ and data ranges. Similar to the previous Fig. \ref{fig27}, the plot highlights that beyond a moment $\qplus$ finite-size corrections no longer follow the canonical expansion Eq. \eqref{e12}.}
	\label{fig28}
\end{figure}

In order to explore the possibility  
of getting good fits with $y$ value 
situated in the interval $y\in[0.75,1.25]$ also at $q\gtrsim \qplus$, we have made further tests. The results for the fitting parameters have been displayed in Fig. \ref{fig28}, where the cases $y=0.75, 1.0$ are compared. As is seen from the goodness of fit, Fig. \ref{fig28} top panel, including higher orders in $L^{-y}$ does not seem to properly describe the finite-size effects in the regime $q\gtrsim \qplus$. 

Taken at face value, the finite-size corrections to the variance appear to change their nature for moments crossing the point $q{\approx}\qplus$. We interpret this observation with an eye on the tail of the distribution function $\mathcal{P}_q(P_q;L)$ discussed in section \ref{ss:tail}. 
The variance $\sigma_q$ as defined in Eq. \eqref{e9} requires the calculation of the second moment of the distribution, which exists only if $\zeta_q\geq 2$. The observation suggests a precise definition of $\zeta_{\qplus}=2$, yielding an estimate  $\qplus\approx 2.7$ based on the results of Fig. \ref{fig22}. This estimate is consistent with the apparent decrease of $y$ that we witness in Figs. \ref{fig27} and \ref{fig28}. We mention that a moment similar to $\qplus$, which is associated with the IPR variance, exists also for the IPR average.  
It is defined as $\zeta_{q_+}=1$; at $q>q_+$ average and typical IPR cease to scale alike with system size. \cite{Evers2008RMP}     

\subsubsection{Estimating $\gamma$ upon including corrections to scaling}
\begin{figure}[t]
	\includegraphics{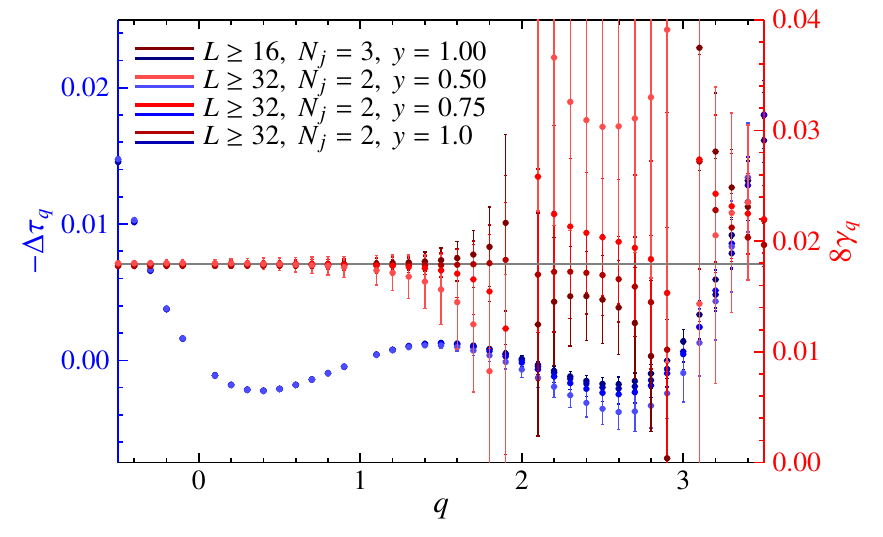}
	\caption{Estimate of the non-parabolic multifractal content $\Delta\tau_q\coloneqq \tau_q-\tau_q^\text{(p)}$  and the corresponding quartic curvature $\gamma_q$ as function of $q$. Fits are based on $\overline{P_q}L^{\tau_q^{(p)}}$ for several fit and parameter ranges. The horizontal line marks $8\gamma=0.01795$. \label{fig18ext}}
\end{figure}
We perform an analysis of finite-size effects following the conventional expansion Eq. (\ref{e8}) for the average IPR $\overline{P_q}$. The main goal is to  quantify deviations from parabolicity, $\delta\tau_q=\delta x_q$, including finite-size corrections. 
Since above analysis suggest a fixed exponent $y\approx 1$ only for moments $q<\qplus$, we compare fits involving
a range $y\in\{0.5, 0.75, 1.0\}$.
Figure \ref{fig18ext} shows the multifractal exponent $\tau_q$ obtained from such fits, represented as deviation from parabolicity $\Delta\tau_q\coloneqq \tau_q-\tau_q^{(p)}$. 
As one would expect based on the analysis of $\sigma_q(L)$ above, fitting gives consistent results for $\Delta \tau_q$ in the regime $q\lesssim 2$. The resulting curvature amounts to $8\gamma\approx 0.0179$ with small statistical error bars that, however, exhibit a significant $q$-dependence. 
We estimate $8\gamma=0.0178\pm 0.0012$ based on the error bars we obtain near $q\approx 0$. 

 With $q$ approaching $\qplus{\approx} 2.7$ from below, the error bars are seen to proliferate dramatically. Importantly, within the error bars the reciprocity relation, $\Delta \tau_q=\Delta \tau_{3-q}$, is seen to be fulfilled in the range of moments investigated,  $-1/2\lesssim q\lesssim 3.5$.

\subsection{Eigenvectors at neighboring energies \label{aC}}
Per sample we calculate three pairs of eigenvectors with eigenvalues nearest to unity, see App. \ref{aA}. The results in the main paper include only one of these  wavefunctions, i.e. the one with eigenvalue closest to unity. We here present a brief analysis of the properties of the other two wavefunctions with eigenvalues next nearest and next-next nearest to unity. 
\begin{figure}[]
	\includegraphics{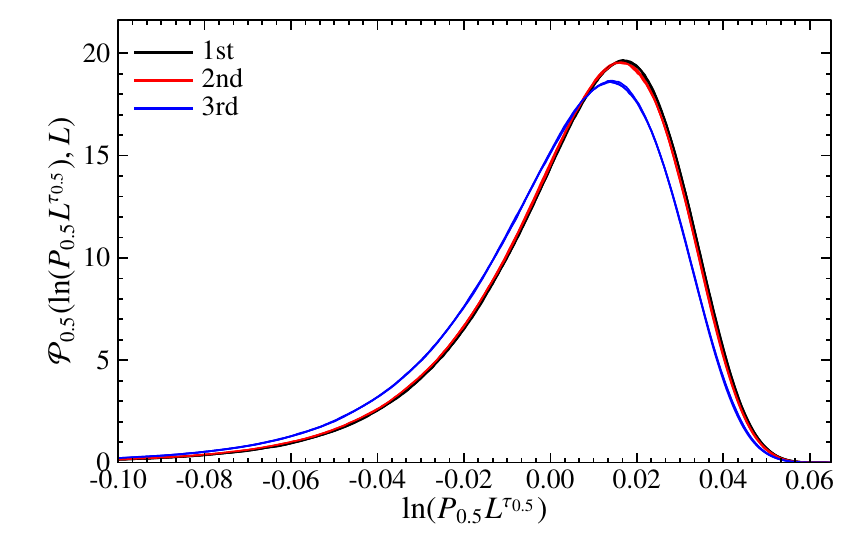}
	\caption{Distribution functions ${\mathcal P}_{0.5}(\ln\left[P_{0.5}(L)L^{\tau_{0.5}}\right];L)$ for $q=0.5$ and $L=512,768,1024$ obtained for three sets of eigenstates. 
	\label{fig30}}
\end{figure}

Figure\ \ref{fig30} shows the scaled IPR distribution function  
${\mathcal P}_{q}(\ln\left[P_{q}(L)L^{\tau_{q}}\right];L)$
at $q=1/2$ for all three eigenstates taken at $8\gamma=0.00178$. 
The excellent data collapse illustrates that $\tau_q$ is the same for all three energies, despite the fact that $\mathcal{P}_q$ is not.  
The collapse is illustrated for a wider range of $q$-values in Fig. \ref{fig31}. 
It displays the auxiliary quantity $\tilde{F}_q(L)$ for the second and third closest eigenstate. Similar to the case of the first eigenvector, Fig.\ \ref{FirstFig19}, also the second and third nearest eigenstates exhibit the hypercollapse. 
However, for these eigenvectors the finite-size effects turn out to be stronger; the collapse is seen to occur only at larger system sizes, i.e. above $L=384$ or $L=512$. 
         
\begin{figure}[t]
    \includegraphics{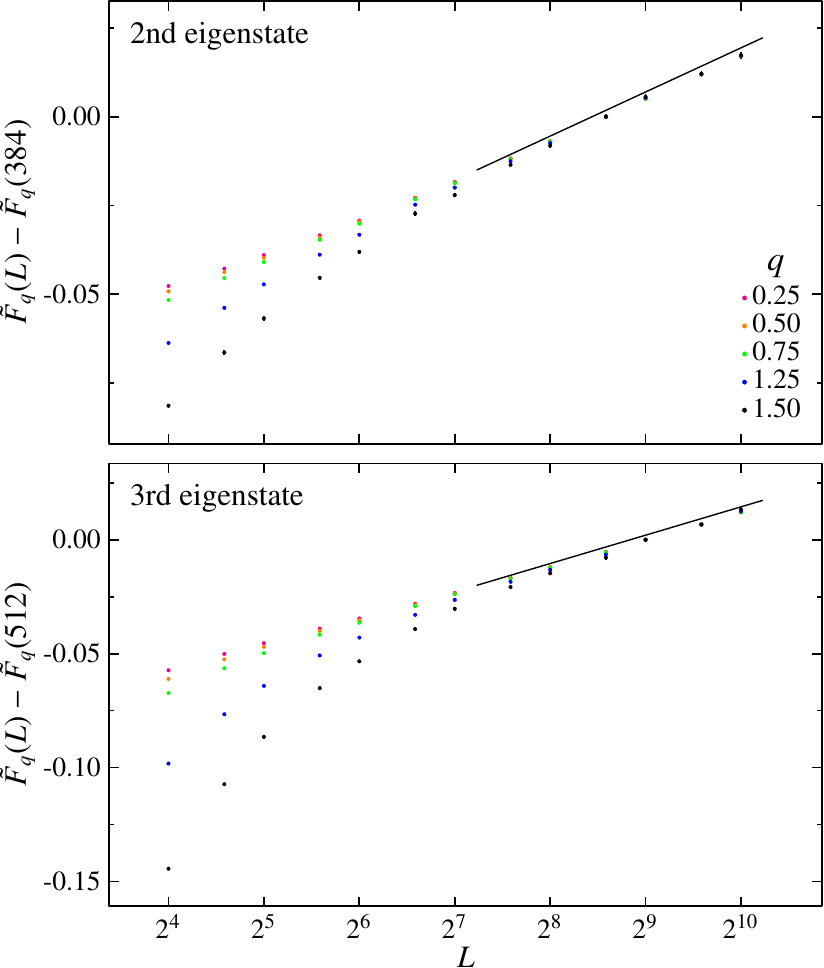}
    \caption{Similar to Fig.\ \ref{FirstFig19}, but for the eigenstates with eigenvalues second nearest (upper panel) and third nearest (lower panel) to unity. The "hyper-collapse" happens at a larger length scale; we changed the reference length $L_0$ to $384$ and $512$ for $2$nd and $3$rd eigenstate, respectively. The slopes of the guiding lines on both panels as well as in Fig.\ \ref{FirstFig19} are equal. \label{fig31}}
\end{figure}

\subsection{IPR dependence on microscopic definition \label{aD}} 
\begin{figure}[]
    \includegraphics[]{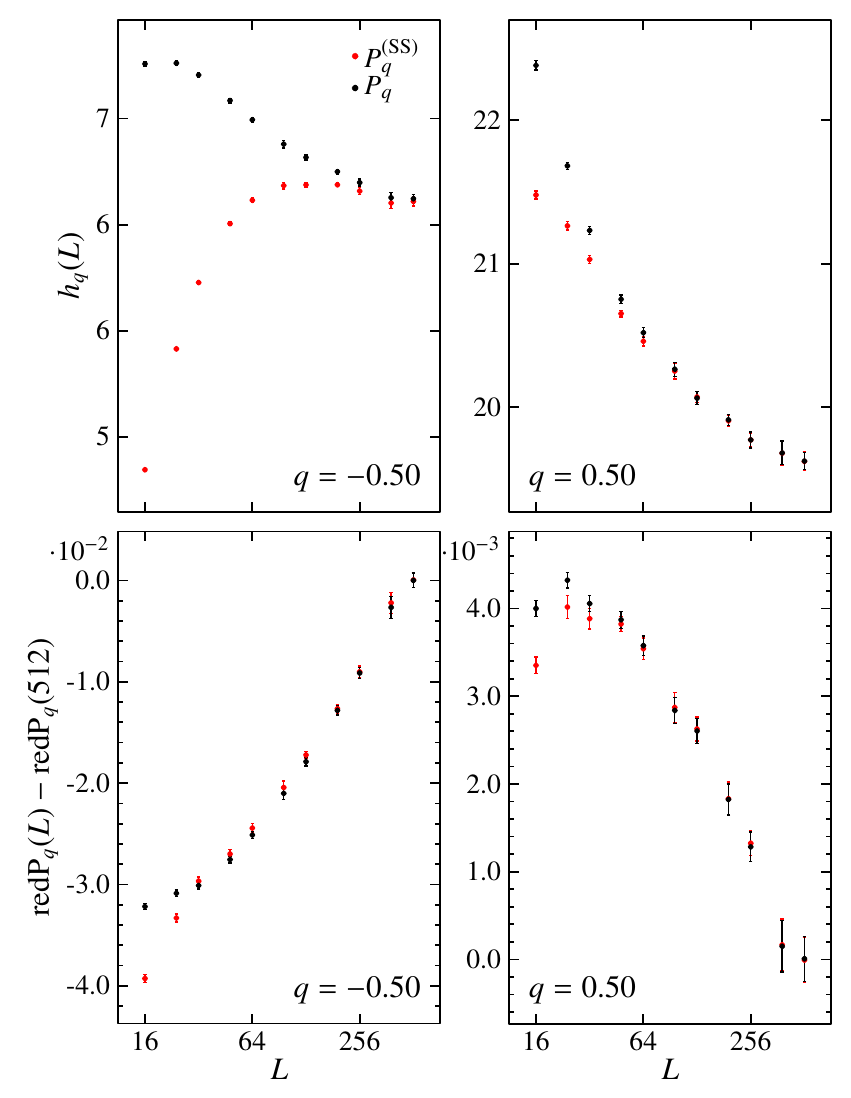}
	\caption{Lower panels show the flow of the height $h_{0.5}(L)$ and maximum position $\text{redP}_{0.5}(L)$ as function of $L$. The later is shown with respect to the position of the largest system, $L=512$. The collapse for $L\geq64$ emphasizes the independence on the microscopic details. 
	\label{fig32}}
\end{figure}
\begin{table}[]
    \begin{tabular}{l|rrrrrrr}
    $L$ & 16 & 24 & 32 & 48 & 64 & 96 & 128	\\ 	\hline
    $\Nsamples(L)$ & 5000 & 5000 & 5000 & 5000 & 5000 & 1081 & 2518\\[2mm]
     $L$ & 192 & 256 & 384 & 512 & 768 & 1024 &\\ \hline
    $\Nsamples(L)$ & 2383 & 1558 & 393 & 582 & - & - &
    \end{tabular}
     \caption{Number of lattice realizations $\Nsamples$ (in units of $1000$) as a function of system size $L$ for the test calculation regarding the microscopic definition. \label{t4}}
\end{table}

By definition the IPR is a sum over space of a local measure, $\mu(\br)$,  taken to the power $q$: $\int d\br\ \mu(\br)^q$. The definition employed in Eq. (\ref{e7}) on the lattice amounts to 
$\mu_l = \sum_\sigma |\psi_{l\sigma}|^2 $. 
An alternative local measure is given by $\mu_l=|\psi_{l\sigma}|^2$, so the statistical properties of the spin density are evaluated for each component, separately. The local density of a given spin direction is more sensitive to rare events as compared to the local number density. We here present a sanity check indicating that the multifractal spectrum is not affected by this difference, at least not in the $q$-window of most interest to us. 

We define the spin-separated IPR
\begin{equation}
   \Pss_q(L) = \sum_l \sum_\sigma |\psi_{l\sigma}|^{2q}. 
\end{equation}
We have as usual 
$\Pss_{q=1}=1$ from normalization, while $\Pss_{q=0}=2L^2$ as opposed to $P_{q=0}=L^2$; the definition of $\tau_q$ is unaffected by this detail. 
Qualitative deviations between the scaling properties of $P_q$ and $\Pss_q$ are expected for $q$ approaching more and more negative values. 

To illustrate similarities and dissimilarities, we have performed a separate study considering systems up to linear system size $L=512$, see Table \ref{t4} for the sample statistics.
The results of these calculations are summarized in Fig. \ref{fig32} for the paradigmatic cases $q=\pm 1/2$. The plot allows us to draw several conclusions. 
(i) At positive $q$, the distributions of  $P_q$ and $\Pss_q$ exhibit a very similar shape, represented by $h_q(L)$ in the top row of Fig. \ref{fig32}. Likewise, the evolution of the distributions with increasing system size is the same, confirming the same set of multifractal indices $\tau_{1/2}$. Corresponding evidence is given in Fig. \ref{fig32}, lower row that shows the flow the peak position of the reduced distributions $\mathcal{P}_q(\ln P_q L^{\tau_q^\text{(p)}})$ 
and $\mathcal{P}_q(\ln \Pss_q L^{\tau_q^\text{(p)}})$: after performing a rigid shift both traces collapse indicating that the critical exponent is the same for both measures. 
(ii) With respect to the critical exponent the situation is seen to be similar at $q=-1/2$, as illustrated in the bottom row of Fig. \ref{fig32}. The form of the distribution functions begins to change shape, however, as clearly displayed in Fig. \ref{fig32} (top row) by $h_{-1/2}(L)$. We take this observation as a precursor for a qualitative deviation of the critical behavior occurring at more negative $q$-values. 

\end{document}